\documentclass[11pt]{article}
\pdfoutput=1
\usepackage{amssymb}
\usepackage{amsmath,array,slashed}
\usepackage{amsthm}
\usepackage{epsfig,graphics}
\usepackage[pdftex]{color}
\usepackage{simplemargins}
\usepackage{graphicx}

\setleftmargin{0.7in}
\setrightmargin{0.7in}

\newcommand{\inp}[1]{\langle #1 \rangle}
\newcommand{\Dl}{\Delta}

\begin{document}

\begin{center}

\hfill  Imperial/TP/2011/LIU/01

\vskip 2 cm
{\Large \bf Next-next-to-extremal Four Point Functions of $\mathcal{N}=4$ $1/2$-BPS Operators  in the AdS/CFT Correspondence \\}
{\vskip 1cm  
Linda I. Uruchurtu\\
Department of Physics \\
The Blackett Laboratory \\
Imperial College \\
London  SW7 2AZ}

\end{center}

\vskip 2 cm
\abstract{Four point functions of general $\mathcal{N}=4$ \ 1/2-BPS primary fields, satisfying the next-next-to-extremality condition $\Delta_{1}+\Delta_{2}+\Delta_{3}-\Delta_{4}=4$ are studied at large $N$ and strong coupling. We apply new techniques to evaluate the effective couplings in supergravity, and confirm that the four derivative couplings arising in the five-dimensional supergravity vanish on-shell.

We then show that the four point amplitude resulting from supergravity naturally splits into a ``free'' and an interactive part which resembles an effective quartic interaction. The precise structure agrees with superconformal symmetry and supports the conjecture formulated by Dolan, Osborn and Nirschl regarding the strongly coupled form of four point correlators of chiral primary operators. We also evaluate the amplitude in large $N$ free field SYM theory and discuss the results in the context of the correspondence.}

\newpage
\section{Introduction}
Correlation functions of chiral primary operators (CPOs) in $\mathcal{N}=4$ super Yang-Mills theory were widely studied during the early days of the AdS/CFT correspondence \cite{Maldacena:1997re, Witten:1998qj, Gubser:1998bc}. Given the non-trivial nature of the celebrated duality, computations were carried out in the supergravity theory (which is weakly coupled) to determine the strongly coupled limit of the corresponding gauge theory correlator, as the infinite tower of KK scalar excitations are dual via the standard AdS/CFT dictionary, to all 1/2-BPS chiral primary operators. Superconformal symmetry already imposes severe constraints on the form of the correlation functions. These operators are known to have protected conformal dimensions and their three point functions having their spatial dependence fixed up to an overall normalisation constant, computed in both supergravity and free field theory for large $N$ and shown to agree \cite{Lee:1998bxa}. 

Four point functions are the first truly dynamical objects, as they are not completely constrained by non-renormalisation theorems and depend on $g$. This is reflected on the presence of operators belonging to long multiplets when they are given an Operator Product Expansion (OPE) interpretation \cite{Arutyunov:2000ku,Dolan:2000ut,Dolan:2001tt,Dolan:2002zh}, with the anomalous dimensions of the operators occurring as a perturbative expansion in $g$.  A four point function can be given by written as a double OPE expansion of the form
 \begin{equation}
\langle \mathcal{O}_i(\vec{x}_1)\mathcal{O}_j(\vec{x}_2)\mathcal{O}_k(\vec{x}_3)\mathcal{O}_z(\vec{x}_4) \rangle = \sum_{\Delta,l} \frac{C_{\mu_1 \cdots \mu_l}(\vec{x}_{12},\partial_{\vec{x}_2})C_{\nu_1 \cdots \nu_l}(\vec{x}_{34},\partial_{\vec{x}_4})}{\vert \vec{x}_{12} \vert^{\Delta_i+\Delta_j-\Delta-l}\vert \vec{x}_{34} \vert^{\Delta_k+\Delta_l-\Delta-l}}\Bigg(\frac{I_{\mu_1\cdots \mu_l \nu_1 \dots \nu_l}(\vec{x}_{24})}{\vert \vec{x}_{24}\vert^{2\Delta}}\Bigg)
\label{OPEdouble}
\end{equation}
Here $C_{\mu_1\cdots \mu_l}$ are the operator algebra structure constants which are fixed by the conformal dimensions and by the three point functions normalisation constants. Hence determination of the spectrum and of all three point normalisation constants will yield the theory solvable as any higher point function can be obtained using the OPE. In the literature, four point functions have been computed for  chiral primary operators belonging to the $[0,p,0]$ representation for $p=2, 3, 4$ \cite{Arutyunov:2000py, Arutyunov:2002fh, Arutyunov:2003ae} and for the mixed case involving two operators in the $[0,2,0]$ and two operators in the $[0,p,0]$ \cite{Berdichevsky:2007xd,Uruchurtu:2008kp}. These operators are dual to Kaluza-Klein supergravity fields $s_p$ with masses $m^2 = p(p-4)$ arising from the reduction of the IIB five form flux and the trace of the ten dimensional metric. The supergravity results generated contributions of order $O(1/N^2)$ from connected diagrams which reproduced the results coming from free field Yang-Mills theory to that order, plus an additional dynamical piece due to interactions.

In general, using the AdS/CFT correspondence, any given higher point function of chiral primary operators will be given in terms of $AdS_{d+1}$ integrals of the form
\begin{equation}
D_{\Delta_1 \Delta_2 \cdots \Delta_n}(\vec{x}_1,\vec{x}_2,\cdots,\vec{x}_n)=\frac{1}{\pi^{d/2}}\int [dz] \prod^{n}_{i=1}\left(\frac{z_0}{z_0^2+(\vec{z}-\vec{x}_i)^2}\right)^{\Delta_i}
\label{genD}
\end{equation}
where $\vec{x}_i$ are the points in the boundary of $AdS_{d+1}$. These $D$-functions transform covariantly under conformal transformations and in the $n=3$ case reduce to the standard form of the three point function. When $n=4$, (\ref{genD}) becomes essentially a quartic vertex and only depends on two conformal invariants $u$, $v$ and any four point function can be shown to reduce to a linear combination of $\bar{D}$-functions.

For the cases studied in the literature, the dynamical contribution obtained from all supergravity four point amplitudes, has been shown to reduce to an expression involving $\bar{D}$-functions\footnote{The relation between $\bar{D}$-functions and $D$-functions can be found in appendix \ref{sec:Dfunc}.} of the form $\bar{D}_{i p+2 j k}$ for various $i,j,k \leq p$. This is precisely what is required so the OPE has no contributions from twist\footnote{The twist is given by $\Delta-l$, where $\Delta$ is the scaling dimension and $l$ is the spin. Long multiplets containing operators with non-vanishing anomalous dimensions have twists $\Delta-l\geq 2p$.} two operators belonging to long multiplets in the strong coupling limit \cite{Dolan:2006ec}. With this guiding principle, Dolan, et.al. conjectured that the dynamical piece of any four point function of single trace $[0,p,0]$ chiral primary operators can be expressed in terms of $\bar{D}$-functions of the form $\bar{D}_{i p+2 j k}$ constrained so that in each channel unitarity bounds on the operators appearing in the OPE are satisfied. 

Given the technical difficulty associated with the supergravity computations, further tests of the previous conjecture have been left on hold, and the availability of the spectrum has limited the scope of examples, as these become overly complicated as soon as more massive states are considered. 


More recently, there there has been a renewed interest in computing correlation functions holographically.  In \cite{Janik:2010gc} Janik, et al. introduced a formalism for using semiclassical methods for computing correlation functions of operators dual to massive classical string states. He successfully computed two point functions of operators dual to classical spinning states. Tseytlin and Buchbinder \cite{Buchbinder:2010vw} went a step further and showed how to define the relevant string states and how to translate them into boundary condiitons for the corresponding semiclassical world surfaces. Later on, Zarembo \cite{Zarembo:2010rr}, Costa et al. \cite{Costa:2010rz} and  Roiban and Tseytlin \cite{Roiban:2010fe} developed methods to evaluate three point functions for the particular case in which two states are semiclassical (i.e. dual to a classical string) and the other is dual to a supergravity mode (i.e. a 1/2-BPS chiral primary operator). Buchbinder and Tseytlin \cite{Buchbinder:2010ek} then considered four point functions with two operators dual to supergravity modes (``light''). All of these calculations have been shown to agree with expectations from free field theory, but an apparent disagreement with the supergravity result in \cite{Uruchurtu:2008kp} was highlighted and the issue remains unresolved.

In this paper, our aim is to take a step further and using previous and new methods in supergravity, attempt to verify the Dolan / Nirschl / Osborn conjecture for a more generic example involving chiral primary operators, while also shedding some light into the apparent disagreement uncovered by Buchbinder and Tseytlin in \cite{Buchbinder:2010ek}. We focus our attention in a next-next-to-extremal process involving two operators of conformal dimension $\Delta=k+2$ with $k>2$, an operator with $\Delta=n-k$ so $n-k \geq 2$ and an operator with $\Delta=n+k$. This example is a more general version of the correlator studied in \cite{Uruchurtu:2008kp}, with the subtlety that all operators here are massive and three different operators are present in the calculation, which amount to the emergence of a wider variety of interactions to be studied. 

After establishing the general structure by restricting its dependence using conformal symmetry, we compute the amplitude in AdS supergravity by evaluating the on-shell action. We then compare the result against the predictions made in the gauge theory side, by means of rewriting the result so it is given in terms of unique dynamical function of the form specified by the conjecture in \cite{Dolan:2006ec}. 

To this end, we have followed the standard procedure for computing four point functions in supergravity. Namely, we use the relevant terms listed in \cite{Arutyunov:1998hf, Lee:1998bxa, Lee:1999pj, Arutyunov:1999en, Arutyunov:1999fb} and evaluate the exchange integrals using the techniques in \cite{D'Hoker:1999ni,Arutyunov:2002fh,Berdichevsky:2007xd}. For the evaluation of the effective vertices coming from the integrals over the $S^5$, we build on top of the methods introduced in \cite{Uruchurtu:2008kp} and use some results from \cite{Osborn:2010sp} to give closed expressions to the sums of products of $SO(6)$ C-tensors in terms of polynomials of $SU(4)$ invariants. Using these, we show that the five-dimensional on-shell lagrangian is of the $\sigma$-model type (i.e. four-derivative terms vanish) and that the quartic lagrangian acquires a simple form after a remarkable simplification. Finally, we show how to rewrite the result for the strongly coupled four point amplitude in terms of a unique dynamical function plus a ``free'' part, emulating the behaviour observed in all other examples in the literature (partial non-renormalisation \cite{Eden:2000bk,Arutyunov:2002fh}).

The plan of this paper is as follows. In section \ref{sec:structure} we discuss the structure of the correlation function we are interested in computing by constraining its form using the symmetries of the theory, and show how it can be re-expressed in terms of conformal ratios and $SU(4)$ invariants. We then turn to free Yang-Mills theory in section \ref{sec:FreeFieldTh} and evaluate the various diagrams in the large $N$ limit. In section \ref{sec:Sugracalc} we setup the supergravity calculation by writing the relevant terms of the lagrangian and evaluating explicitly the effective couplings and exchange integrals (technical details are included in appendices \ref{sec:integrationsphere}, \ref{sec:QuarticInt}, \ref{sec:exintegrals}). Section 5 discusses the results and shows how the supergravity amplitude is split into a ``free'' and an interactive piece which is of the same form specified by \cite{Dolan:2006ec}. We also comment on the issues pertaining to the free part and the apparent disagreement with the classical string theory results in \cite{Buchbinder:2010ek}. Section 6 summarises our findings and suggests some possible research avenues to be pursued in the future.

\section{Structure of the Correlation Function}
\label{sec:structure}
We are interested in computing the four point function of $1/2$-BPS superconformal primaries of $\mathcal{N}=4$ supersymmetric Yang-Mills theory. The structure of this correlator is constrained by the symmetries of the theory, comprising $R$-symmetry and crossing symmetry. The canonically normalised operators \cite{Lee:1998bxa}  with conformal dimension $\Delta=p$ are given by
\begin{equation}
\mathcal{O}_p^{r}(\vec{x})=\frac{1}{\sqrt{p N^p}}C_{i_1\cdots i_p}^{r}\mathrm{tr}(\varphi_r^{i_1}(\vec{x})\cdots \varphi_r^{i_p}(\vec{x}))
\label{normCPOk}
\end{equation}
where $C_{i_1\cdots i_p}^r$ is a totally symmetric traceless $SO(6)$ tensor of rank $p$ and the index $r$ runs over a basis of a representation of $SO(6)$. The four point function we wish to study has the form
\begin{equation}
\langle \mathcal{O}^{1}_{k+2}(\vec{x}_1) \mathcal{O}^{2}_{k+2}(\vec{x}_2) \mathcal{O}^{3}_{n-k}(\vec{x}_3) \mathcal{O}^{4}_{n+k}(\vec{x}_4)\rangle
\label{diffweightprocess}
\end{equation}
which is a next-next-to-extremal process\footnote{Recall that the extremality $E$ of a four point function is given by $\Delta_1+\Delta_2+\Delta_3-\Delta_4=2E$ so $E=2$ results in a next-next-to-extremal process.}, given that $\Delta_1+\Delta_2+\Delta_3-\Delta_4=4$, which will constrain the selection rules as to allow us to perform the calculation. Notice that we've chosen the first two operators in the correlator to have the same conformal weight to simplify the computations. The content of the OPE's is given by operators in the representations arising in the tensor product of the $SU(4)$ representations $[0,k+2,0]$, $[0,n-k,0]$ and $[0,n+k,0]$. This is, all $SU(4)$ representations that are in the tensor product of $[0,k+2,0]\otimes [0,k+2,0]$ and $[0,n-k,0]\otimes[0,n+k,0]$ for the $s$-channel 
\begin{align}
\left( [0,k+2,0]\otimes [0,k+2,0] \right)& \cap \left( [0,n-k,0]\otimes [0,n+k,0] \right) 
\nonumber \\
&=[0,2k,0]\oplus [0,2k+2,0]\oplus [0,2k+4,0]  \oplus [1,2k,1] \oplus [1,2k+2,1]\oplus [2,2k,2] 
\label{4pdiffweightreps}
\end{align}
and those in the tensor product of $[0,k+2,0]\otimes [0,n-k,0]$ and $[0,k+2,0]\otimes[0,n+k,0]$ for the $t$-channel
\begin{align}
\left( [0,k+2,0]\otimes [0,n-k,0] \right) &\cap \left( [0,k+2,0]\otimes [0,n+k,0] \right)  
\nonumber \\
&=[0,n-2,0]\oplus [0,n,0]\oplus [0,n+2,0]  \oplus [1,n-2,1] \oplus [1,n,1]\oplus [2,n-2,2] 
\end{align}
Here we used
\begin{equation}
[0,p_1,0]\otimes[0,p_2,0]=\sum_{j=0}^{p_1}\sum_{l=0}^{p_1-j}[j,p_2-p_1+2l,j] 
\end{equation}
where $p_{1 }\leq p_{2}$. All the OPE channels with $j=0,1$ contain only short and semishort operators. We now follow the ideas and methods in \cite{Arutyunov:2002fh}. An appropriate basis to study the content of a four point function is given by the \emph{propagator basis} arising in free field theory. Recall that the propagator for scalar fields is given by
\begin{equation}
\label{scalarpropYM}
\langle \varphi^{i}(\vec{x}_1)\varphi^j(\vec{x}_2)\rangle=\frac{\delta^{ij}}{|\vec{x}_{12}|^2}
\end{equation}
Let us introduce the harmonic (complex) variables $t_{i}$ satisfying the following constraints
\begin{equation}
t_it_i=0 \qquad \qquad t_i\bar{t}_i=1
\end{equation}
These variables parametrise the coset $SO(6)/SO(2)\times SO(4)$ so that under an $SO(6)$ transformation, the highest weight vector representation transforms as $t_{i_1}\cdots t_{i_p}$, so projections onto representations $[0,p,0]$ can be achieved by writing
\begin{equation}
\mathcal{O}_{p}(\vec{x},t)=\frac{1}{\sqrt{pN^p}}t_{i_1}\cdots t_{i_p}\mathrm{tr}(\varphi^{i_1}(\vec{x})\cdots \varphi^{i_p}(\vec{x}))
\label{normgaugeth}
\end{equation}
with $p$ denoting the highest weight of the representation $[0,p,0]$. 
Using these notation, we can rewrite (\ref{scalarpropYM}) as
\begin{equation}
t_i t_j\langle \varphi^i(\vec{x}_1)\varphi^j(\vec{x}_2) \rangle =
\frac{ {t_1}\cdot{t_2}}{|\vec{x}_{12}|^2}
\end{equation}

\begin{figure*}[h!]
\begin{center}
\resizebox{2.8in}{4.3in}{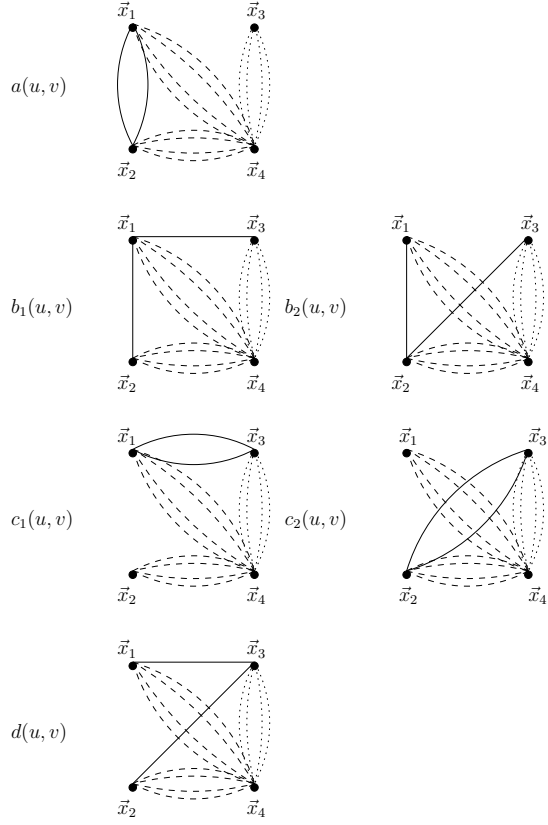} 
\end{center}
\caption{Diagramatic representation for the free contribution to the process $\langle \mathcal{O}_{k+2}\mathcal{O}_{k+2}\mathcal{O}_{n-k}\mathcal{O}_{n+k}\rangle$. The graphs are arranged in four equivalence classes.}
\label{colourbasis}
\end{figure*}

We can now construct four point functions by connecting pairs of points by propagators. For the case in hand, the amplitude will be expressed in terms of the propagator basis for (\ref{4pdiffweightreps}), determined from six graphs belonging to four equivalence classes, as depicted in figure \ref{colourbasis}. 
Each of the propagator structures can be multiplied by an arbitrary function of the conformally invariant ratios $u$ and $v$
\begin{equation}
u=\frac{|\vec{x}_{12}|^2|\vec{x}_{34}|^2}{|\vec{x}_{13}|^2|\vec{x}_{24}|^2} \qquad \qquad
v=\frac{|\vec{x}_{14}|^2|\vec{x}_{23}|^2}{|\vec{x}_{13}|^2|\vec{x}_{24}|^2} 
\label{crossradii}
\end{equation}
where $\vec{x}_{ij}=\vec{x}_i-\vec{x}_j$, so the most general four point amplitude with the required transformation properties is given by
\begin{equation}
\begin{split}
\langle \mathcal{O}_{k+2}(\vec{x}_1,t_1) & \mathcal{O}_{k+2}(\vec{x}_2,t_2)  \mathcal{O}_{n-k}(\vec{x}_3,t_3)\mathcal{O}_{n+k}(\vec{x}_4,t_4)\rangle
\nonumber \\
& = a(u,v)\left(\frac{t_{1}\cdot t_{2}}{|\vec{x}_{12}|^{2}}\right)^{2} \left(\frac{t_{1}\cdot t_{4}}{|\vec{x}_{14}|^{2}}\right)^{k} \left(\frac{t_{2}\cdot t_{4}}{|\vec{x}_{24}|^{2}}\right)^{k} 
\left(\frac{t_{3}\cdot t_{4}}{|\vec{x}_{34}|^{2}}\right)^{n-k}  
\nonumber \\
&+b_1(u,v)\left(\frac{t_{1}\cdot t_{2}}{|\vec{x}_{12}|^{2}}\right)\left(\frac{t_{1}\cdot t_{3}}{|\vec{x}_{13}|^{2}}\right) \left(\frac{t_{1}\cdot t_{4}}{|\vec{x}_{14}|^{2}}\right)^{k} \left(\frac{t_{2}\cdot t_{4}}{|\vec{x}_{24}|^{2}}\right)^{k+1} 
\left(\frac{t_{3}\cdot t_{4}}{|\vec{x}_{34}|^{2}}\right)^{n-k-1}  
\nonumber \\
&+b_2(u,v)\left(\frac{t_{1}\cdot t_{2}}{|\vec{x}_{12}|^{2}}\right)\left(\frac{t_{2}\cdot t_{3}}{|\vec{x}_{23}|^{2}}\right) \left(\frac{t_{1}\cdot t_{4}}{|\vec{x}_{14}|^{2}}\right)^{k+1} \left(\frac{t_{2}\cdot t_{4}}{|\vec{x}_{24}|^{2}}\right)^{k} 
\left(\frac{t_{3}\cdot t_{4}}{|\vec{x}_{34}|^{2}}\right)^{n-k-1}
\nonumber \\
&+c_1(u,v)\left(\frac{t_{1}\cdot t_{3}}{|\vec{x}_{13}|^{2}}\right)^{2}\left(\frac{t_{1}\cdot t_{4}}{|\vec{x}_{14}|^{2}}\right)^{k} \left(\frac{t_{2}\cdot t_{4}}{|\vec{x}_{24}|^{2}}\right)^{k+2} 
\left(\frac{t_{3}\cdot t_{4}}{|\vec{x}_{34}|^{2}}\right)^{n-k-2} 
\nonumber \\
&+c_2(u,v)\left(\frac{t_{1}\cdot t_{4}}{|\vec{x}_{14}|^{2}}\right)^{k+2}\left(\frac{t_{2}\cdot t_{3}}{|\vec{x}_{23}|^{2}}\right)^{2} \left(\frac{t_{2}\cdot t_{4}}{|\vec{x}_{24}|^{2}}\right)^{k} 
\left(\frac{t_{3}\cdot t_{4}}{|\vec{x}_{34}|^{2}}\right)^{n-k-2}
\nonumber \\
&+d(u,v)\left(\frac{t_{1}\cdot t_{3}}{|\vec{x}_{13}|^{2}}\right)\left(\frac{t_{2}\cdot t_{4}}{|\vec{x}_{24}|^{2}}\right)^{k+1} \left(\frac{t_{2}\cdot t_{3}}{|\vec{x}_{23}|^{2}}\right) 
\left(\frac{t_{1}\cdot t_{4}}{|\vec{x}_{14}|^{2}}\right)^{k+1}\left(\frac{t_{3}\cdot t_{4}}{|\vec{x}_{34}|^{2}}\right)^{n-k-2}
\label{structure4p}
\end{split}
\end{equation}

Let us introduce the $SU(4)$ invariants which are homogeneous of degree zero and are defined by
\begin{equation}
\sigma = \frac{t_{1}\cdot t_{3} \, t_{2}\cdot t_{4}}{t_{1}\cdot t_{2} \, t_{3} \cdot t_{4}} \qquad \qquad \tau= \frac{t_{1}\cdot t_{4} \, t_{2}\cdot t_{3}}{t_{1}\cdot t_{2}\, t_{3} \cdot t_{4}} 
\label{sigmatau}
\end{equation}
with these definitions we can recast (\ref{structure4p}) in the following form
\begin{align}
\langle  \mathcal{O}_{k+2}(\vec{x}_1,t_1) \mathcal{O}_{k+2}(\vec{x}_2,t_2) \mathcal{O}_{n-k}(\vec{x}_3,t_3)\mathcal{O}_{n+k}(\vec{x}_4,t_4)\rangle
&
\nonumber \\
=\left(\frac{t_{1}\cdot t_{2}}{|\vec{x}_{12}|^{2}}\right)^{2} \left(\frac{t_{1}\cdot t_{4}}{|\vec{x}_{14}|^{2}}\right)^{k} &\left(\frac{t_{2}\cdot t_{4}}{|\vec{x}_{24}|^{2}}\right)^{k} 
\left(\frac{t_{3}\cdot t_{4}}{|\vec{x}_{34}|^{2}}\right)^{n-k}  \mathcal{G}(u,v;\sigma,\tau)
\label{structure4pCten}
\end{align}
where $\mathcal{G}(u,v;\sigma,\tau)$ is given as a polynomial in $\sigma$ and $\tau$ 
\begin{equation}
 \mathcal{G}(u,v;\sigma,\tau)= a(u,v)+b_1(u,v) u \sigma+b_2(u,v)\frac{u}{v}\tau+c_1(u,v)u^2\sigma^2+c_2(u,v)\frac{u^{2}}{v^2}\tau^2+d(u,v)\frac{u^2}{v}\sigma\tau
 \label{gpoly}
\end{equation}
Crossing symmetry imposes restrictions on $\mathcal{G}(u,v;\sigma,\tau)$. Under exchange of $1\leftrightarrow 2$
\begin{equation*}
(u,v) \rightarrow \left(\frac{u}{v},\frac{1}{v}\right) \qquad \qquad \sigma \leftrightarrow \tau
\end{equation*}
so that
\begin{equation}
\mathcal{G}\left(u,v;\sigma, \tau \right)=\mathcal{G}\left(\frac{u}{v},\frac{1}{v};\tau,\sigma\right)
\label{crosssym4}
\end{equation}
Hence, the number of coefficient functions is reduced to four as
\begin{align}
a(u,v)&=a(u/v,1/v) \nonumber \\
b_2(u,v)&=b_1(u/v,1/v) \nonumber \\
c_2(u,v)&=c_1(u/v,1/v) \nonumber \\
d(u,v)&=d(u/v,1/v)
\end{align}
When $n=2k+2$ there is an additional crossing symmetry $1\leftrightarrow 3$, such that
\begin{align}
\mathcal{G}(u,v;\sigma,\tau)=\left( \tau \frac{u}{v} \right)^2 \mathcal{G}\left(v,u,\frac{\sigma}{\tau},\frac{1}{\tau}\right)
\label{extrasym}
\end{align}
and there are only two independent coefficient functions left.

Ward identities and dynamical considerations force $\mathcal{G}(u,v;\sigma,\tau)$ to split into two distinct pieces, the first related to the result obtained for free fields and the second containing all the contributions coming from non-trivial dynamics to the four point function. Namely,
\begin{equation}
\mathcal{G}(u,v;\sigma,\tau)=\mathcal{G}_0(u,v;\sigma,\tau)+s(u,v;\sigma,\tau)\mathcal{H}_I(u,v;\sigma,\tau)
\label{fullamp}
\end{equation}
where
\begin{equation}
s(u,v;\sigma,\tau)=v+\sigma^2 uv+\tau^2 u +\sigma v (v-u-1) +\tau (1-u-v)+\sigma\tau (u-v-1)
\label{spoly}
\end{equation}
It is easy to see that (\ref{spoly}) transforms under crossing symmetry as $s(u,v;\sigma,\tau)=v^{2}s(u/v,1/v,\tau,\sigma)$. This together with (\ref{crosssym4}) necessarily implies
\begin{align}
\mathcal{H}_I(u,v; \sigma, \tau)=\frac{1}{v^2}\mathcal{H}_I\left( \frac{u}{v},\frac{1}{v}; \tau,\sigma \right)
\label{hcross4}
\end{align}
\section{Free Field Theory at large $N$}
\label{sec:FreeFieldTh}
%
It is possible to calculate the leading large $N$ behaviour for the free field contributions to the four point function. Our calculations will be very similar to those presented in \cite{Dolan:2006ec}. We start by introducing the $U(N)$ generators $\{ T_a \}$, $a=1\cdots N^2$, which are $N \times N$ matrices satisfying
\begin{equation}
[T_a, T_b]=i f_{abc} T_c \qquad \mathrm{tr}(T_a T_b)=\frac{1}{2}\delta_{ab} \qquad \mathrm{tr}(T_a A)\mathrm{tr}( T_a B)=\frac{1}{2}\mathrm{tr}(AB) \qquad
T_a T_a = \frac{1}{2}N \mathbb{I}
\end{equation}
Introducing the basic two-point function of adjoint scalar fields $X=X_a T_a$
\begin{equation}
\langle X_a X_b \rangle = 2\delta_{ab}
\end{equation}
so single trace primary operators belonging to the $[0,p,0]$ representation of $SU(4)$ correspond to $\mathrm{tr}(X^p)$. Consequently, the two-point function of chiral primary operators of equal conformal dimension becomes
\begin{align}
\langle \mathrm{tr}(X^p)\mathrm{tr}(X^p)\rangle&=2^p p! \  \mathrm{tr}(T_{(a_1} \cdots T_{a_p)})\ \mathrm{tr}(T_{(a_1} \cdots T_{a_p)}) \nonumber \\
&\simeq 2^p p \ \mathrm{tr}(T_{a_1} \cdots T_{a_p})\ \mathrm{tr}(T_{a_p} \cdots T_{a_1}) \nonumber \\
&=2^{p-1}p \ \mathrm{tr}(T_{a_1} \cdots T_{a_{p-1}}T_{a_{p-1}}\cdots T_{a_1} )=pN^p
\label{2pnonorm}
\end{align}
where sub-leading terms in the large $N$ limit have been dropped. 

Three point functions can be evaluated analogously, taking into account the relevant symmetry factors
\begin{equation}
\langle  \mathrm{tr}(X^{p_1})\mathrm{tr}(X^{p_2})\mathrm{tr}(X^{p_3}) \rangle = p_1 p_2  p_3 \ N^{\frac{1}{2}(p_1+p_2+p_3)-1}
\end{equation}
In the extremal case, $p_1+p_2=p_3$ and in the large $N$ limit the result is given by $p_1p_2p_3 N^{p_{3}-1}$. We are now ready to evaluate the large $N$ limit of the four point function of interest. There are six possible ways of contracting the chiral primary operators (see figure \ref{colourbasis}). 
\begin{figure*}[!t]
\begin{center}
\resizebox{1.2in}{1.2in}{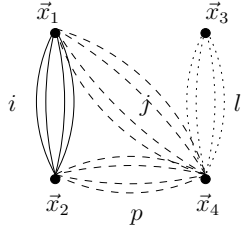} 
\end{center}
\caption{In the limit in which $k\rightarrow 0$, this diagram becomes the disconnected contribution to the four point function.}
\label{diagone2}
\end{figure*}
Consider the first diagram and let $i=2$, $j=p=k$ and $l=n-k$ (see figure \ref{diagone2}). With appropriate symmetry factors one has:
\begin{align}
\langle  \mathrm{tr}(X^{k+2})\mathrm{tr}(X^{k+2})\mathrm{tr}(X^{n-k}) \mathrm{tr}(X^{n+k}) \rangle_a
\nonumber \\
 = \frac{2^{n+k+2}(k+2)!^2(n-k)!(n+k)!}{(k!)^2(n-k)!2!}&\mathrm{tr}(T_{(a_1}T_{a_2}T_{b_1}\cdots T_{b_k)}) \mathrm{tr}(T_{(a_1}T_{a_2}T_{c_1}\cdots T_{c_k)}) \nonumber \\
\times &\mathrm{tr}(T_{(d_1}\cdots T_{d_{n-k})}) \mathrm{tr}(T_{(b_1}\cdots T_{b_k}T_{c_1}\cdots T_{c_k}T_{d_1}\cdots T_{d_{n-k})})
\label{simp1}
\end{align}
Note that
\begin{align}
(n+k)! \ \mathrm{tr}(T_{(b_1}\cdots T_{b_k}T_{c_1}\cdots T_{c_k}& T_{d_1}\cdots T_{d_{n-k})})  
\mathrm{tr}(T_{(d_1}\cdots T_{d_{n-k})})  
\nonumber \\
&=2^{-(n-k)}N^{n-k-1}(n-k)(n+k)(2k)! \ \mathrm{tr}\left(T_{(b_1}\cdots T_{b_j}T_{c_1}\cdots T_{c_k)}\right) \nonumber
\end{align}
substituting this result in (\ref{simp1}) gives:
\begin{align}
2^{2k+2}\frac{(k+2)!^2 (n-k) (n+k) (2k)!}{i! \ j! \ p!}\mathrm{tr}(T_{(a_1}T_{a_2}T_{b_1}\cdots T_{b_k)}) \mathrm{tr}(T_{(a_1}T_{a_2}T_{c_1}\cdots T_{c_k)})\mathrm{tr}\left(T_{(b_1}\cdots T_{b_j}T_{c_1}\cdots T_{c_k)}\right) \nonumber
\end{align}
which looks exactly like a three point function. Proceeding as before and substituting the values of $i$, $j$ and $p$, the final result reads:
\begin{align}
2^{2k+2}\cdot 2^{-\frac{1}{2}(2k+4+j+p)}N^{\frac{1}{2}(2k+4+j+p)-1}(k+2)^2(2k)=N^{n+k}2k(k+2)^2(n-k)(n+k)
\end{align}
since $\frac{1}{2}(2k+4+j+p)=2k+2$. 
\begin{figure*}[!t]
\begin{center}
\resizebox{1.2in}{1.2in}{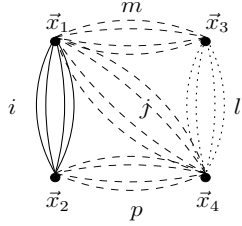} 
\end{center}
\caption{Second diagram}
\label{diagone3}
\end{figure*}
Let us now evaluate the second diagram (see \ref{diagone3}). In this case $i=m=1$, $j=k$, $p=k+1$ and $l=n-k-1$ so the contribution to the four point function from this diagram is given by
\begin{align}
\langle  \mathrm{tr}(X^{k+2})\mathrm{tr}(X^{k+2})\mathrm{tr}(X^{n-k}) \mathrm{tr}(X^{n+k}) \rangle_{b_1}
\nonumber \\
 = \frac{2^{n+k+2}(k+2)!^2(n-k)! \ (n+k)!}{i! \ j! \ p!\ l! \ m!}&\mathrm{tr}(T_{(a_1}T_{b_1}\cdots T_{b_k}T_{e_1)})\  \mathrm{tr}(T_{(a_1}T_{c_1}\cdots T_{c_{k+1})}) \nonumber \\
\times &\mathrm{tr}(T_{(d_1}\cdots T_{d_{n-k-1}}T_{e_1)}) \ \mathrm{tr}(T_{(b_1}\cdots T_{b_k}T_{c_1}\cdots T_{c_{k+1}}T_{d_1}\cdots T_{d_{n-k-1})})
\nonumber
\end{align}
It is useful to record
\begin{align}
(k+2)! \ (n-k)! \ &\mathrm{tr}(T_{(a_1}T_{b_1}\cdots T_{b_k}T_{e_1)})\ \mathrm{tr}(T_{(d_1}\cdots T_{d_{n-k-1}}T_{e_1)})
\nonumber \\
&=\frac{1}{2}(k+2)(n-k)(i+j)! \ l! \ \mathrm{tr}(T_{(a_1}T_{b_1}\cdots T_{b_k)}T_{(d_1}\cdots T_{d_{n-k-1})})
\nonumber \\
\frac{(k+2)! \ (n+k)!}{p!} \ &\mathrm{tr}(T_{(a_1}T_{c_1}\cdots T_{c_{k+1)}})\ \mathrm{tr}(T_{(b_1}\cdots T_{b_{k}}T_{c_1}\cdots T_{c_{k+1}}T_{d_1}\cdots T_{d_{n-k-1})})
\nonumber \\
&=2^{-k-1}N^{p}(k+2)(n+k)(j+l)!  \ \mathrm{tr}(T_{(b_1}\cdots T_{b_k}T_{d_1}\cdots T_{d_{n-k-1})}T_{a_1}) \nonumber
\end{align}
substituting back and simplifying 
\begin{align}
&2^{n}N^{k}(k+2)^2(n-k)(n+k)\frac{(i+j)! \ (j+l)!}{j!}\mathrm{tr}(T_{(a_1}T_{b_1}\cdots T_{b_k)}T_{(d_1}\cdots T_{d_{n-k-1})})
\nonumber \\
&\hspace{64mm} \times \mathrm{tr}(T_{(b_1}\cdots T_{b_k}T_{d_1}\cdots T_{d_{n-k-1})}T_{a_1})\nonumber\\
&=2^{k+1}N^{n-2}(k+2)^2(n-k)(n+k)(i+j)! \  (j+1)! \ \mathrm{tr}(T_{(a_1}T_{b_1}\cdots T_{b_k)}T_{(b_1}\cdots T_{b_{k})}T_{a_1})
\nonumber \\
&=N^{n+k}(k+2)^2(n-k)(n+k)(k+1)
\end{align}
where we used that $i=1$ and $j=k$. Evaluation of the third diagram gives the same result, as expected from crossing symmetry. The remaining diagrams can be evaluated analogously. We summarise our results below:
\begin{align}
\langle  \mathrm{tr}(X^{k+2})\mathrm{tr}(X^{k+2})\mathrm{tr}(X^{n-k}) \mathrm{tr}(X^{n+k}) \rangle_{a}&=N^{n+k}2k(k+2)^2(n-k)(n+k) \nonumber \\
\langle  \mathrm{tr}(X^{k+2})\mathrm{tr}(X^{k+2})\mathrm{tr}(X^{n-k}) \mathrm{tr}(X^{n+k}) \rangle_{b_1,b_2}&=N^{n+k}(k+2)^2(n-k)(n+k)(k+1) \nonumber \\
\langle  \mathrm{tr}(X^{k+2})\mathrm{tr}(X^{k+2})\mathrm{tr}(X^{n-k}) \mathrm{tr}(X^{n+k}) \rangle_{c_1,c_2}&=N^{n+k}(k+2)^2(n-k)(n+k)(n-2) \nonumber \\
\langle  \mathrm{tr}(X^{k+2})\mathrm{tr}(X^{k+2})\mathrm{tr}(X^{n-k}) \mathrm{tr}(X^{n+k}) \rangle_{d_1}&=N^{n+k}(k+2)^2(n-k)(n+k)(n-k-1)
\end{align}
Using the normalisation in (\ref{normgaugeth}), the two point function of chiral primary operators (\ref{2pnonorm}) becomes
\begin{equation}
\langle \mathcal{O}_p(\vec{x}_1,t_1)\mathcal{O}_p(\vec{x}_2,t_2)\rangle=\left(\frac{t_1 \cdot t_2}{\vert \vec{x}_{12}\vert^2}\right)^p
\end{equation}
and using (\ref{structure4pCten}) we can now write down the large $N$ result of $\mathcal{G}_0(u,v;\sigma,\tau)$. This is,
\begin{align}
\mathcal{G}_0(u,v;\sigma,\tau)=\frac{1}{N^2}\sqrt{(k+2)^2(n-k)(n+k)}&\Big\{ 2k+(k+1)\Big(\sigma u + \tau \frac{u}{v}\Big)+(n-2)\Big(\sigma^2+\tau^2\frac{u^2}{v^2}\Big)
\nonumber \\
&+(n-k-1)\sigma\tau \frac{u^2}{v}\Big\}
\label{freeresultamp}
\end{align}
Notice that in the limit in which $n=2k+2$, 
\begin{align}
\mathcal{G}_0(u,v;\sigma,\tau)=\frac{1}{N^2}\sqrt{(k+2)^3(3k+2)}&\Big\{ 2k\Big(1+\sigma^2 + \tau^2 \frac{u^2}{v^2}\Big)+(k+1)\Big(\sigma u+\tau \frac{u}{v}+\sigma\tau \frac{u^2}{v}\Big) \Big\}
\end{align}
which is consistent with (\ref{extrasym}).
\section{Supergravity Calculation}
\label{sec:Sugracalc}
The AdS/CFT correspondence relates $\mathcal{N} =4$ SYM theory and IIB string theory in $AdS_{5} \times S^{5}$. In the limit in which $N$ and $\lambda \gg 1$, one can set up a precise relation between correlation functions of single trace operators in the planar limit and correlation functions of the corresponding supergravity states. The statement between operators in the gauge theory and fields in the bulk was established in  \cite{Witten:1998qj,Gubser:1998bc} and refined in  \cite{Mueck:1999kk,Mueck:1999gi,Bena:1999jv}. The proposition is
 \begin{equation}
\Big\langle \exp\{ \int d^{4}x \phi_{0}(\vec{x})\mathcal{O}(\vec{x})\}\Big\rangle_{CFT}=\exp \{ -S_{IIB}[\phi_{0}(\vec{x})] \}
\label{prescriptionadscft}
\end{equation}
On the left hand side of (\ref{prescriptionadscft}) the field $\phi_0(\vec{x})$, which stands for the boundary value of the bulk field $\phi(z_0,\vec{x})$, is a source for the operator $\mathcal{O}(\vec{x})$, and the expectation value is computed by expanding the exponential and evaluating the correlation functions in the field theory. On the right hand side, one has the generating functional encompassing all dynamical processes of IIB strings on $AdS_5\times S^{5}$. In the supergravity approximation, $S_{IIB}$ is just the type IIB supergravity action on $AdS_5\times S^5$, and it is assumed here that all the bulk fields $\phi(z_0,\vec{x})$ have appropriate boundary behaviour so they source the YM operators on the left hand side. Hence in practice, one first finds the boundary data for the corresponding gravitational fields and then computes correlation functions as a function of these values (on-shell), by functional differentiation.

Given that we are interested in computing correlation functions of superconformal primaries, we first need to identify the bulk fields whose value in the boundary serve as sources.  From looking at the representations, we see that the fields dual to superconformal primaries are obtained from mixtures of modes from the graviton and the five form on the $S^5$ \cite{Kim:1985ez} and are denoted as $s_k^I$, with $I$ running over the basis of the corresponding $SO(6)$ irrep. with Dynkin labels $[0,k,0]$. The four point function can then be determined from the expression
\begin{equation}
\langle \mathcal{O}^{1}_{k_1}(\vec{x}_{1})\mathcal{O}^{2}_{k_2}(\vec{x}_{2})\mathcal{O}^{3}_{k_3}(\vec{x}_{3})\mathcal{O}^{4}_{k_4}(\vec{x}_{4})\rangle=
\frac{\delta}{\delta s_{k_1}^{I_1}(\vec{x}_{1})}\frac{\delta}{\delta s_{k_2}^{I_2}(\vec{x}_{2})}
\frac{\delta}{\delta s_{k_3}^{I_3}(\vec{x}_{3})}\frac{\delta}{\delta s_{k_4}^{I_4}(\vec{x}_{4})}(-S_{IIB})
\end{equation}
\subsection{On-Shell Lagrangian}
\label{subsec:onshelllag}
We are interested in computing the process
\begin{equation}
\langle \mathcal{O}^{1}_{k+2}(\vec{x}_{1})\mathcal{O}^{2}_{k+2}(\vec{x}_{2})\mathcal{O}^{3}_{n-k}(\vec{x}_{3})\mathcal{O}^{4}_{n+k}(\vec{x}_{4})\rangle
\label{4pfunction2do}
\end{equation}
in strongly coupled $\mathcal{N}=4$ SYM theory, using the supergravity approximation. This process satisfies the next-next-to-extremal condition $k_{1}+k_{2}+k_{3}-k_{4}=4$ and we will consider the case in which $n\geq k+2$. The prescription (\ref{prescriptionadscft}) indicates that we need to evaluate the on-shell value of the five-dimensional effective action of compactified type IIB supergravity on $AdS_5\times S^5$. We write this action as
\begin{equation}
S=\frac{N^2}{8\pi^2}\int [dz] \Big(\mathcal{L}_2+\mathcal{L}_3+\mathcal{L}_4\Big)
\end{equation}
which involves the sum of quadratic, cubic and quartic terms. The normalisation of the action can be derived from expressing the ten dimensional gravitational coupling as $2\kappa_{10}^2=(2\pi)^7g_s^2\alpha'^4$ and using the volume of $S^5$ to get the five dimensional gravitational coupling
\begin{equation}
\label{normaction5d}
\frac{1}{2\kappa_{5}^2}=\frac{\mathrm{Vol}(S^5)}{2\kappa_{10}^2}=\frac{N^2}{8\pi^2l^3}
\end{equation}
with $l$ being the $AdS_5$ radius, which will be set to one. The relevant quadratic terms \cite{Kim:1985ez,Arutyunov:1998hf} read
\begin{align}
\mathcal{L}_{2}=&\frac{1}{4}(D_{\mu}{s_{k+2}}D^{\mu}{s_{k+2}}+(k+2)(k-2){s_{k+2}}{s_{k+2}}) \nonumber \\
               +&\frac{1}{4}(D_{\mu}{s_{n-k}}D^{\mu}{s_{n-k}}+(n-k)(n-k-4){s_{n-k}}{s_{n-k}})
               \nonumber \\
               +&\frac{1}{4}(D_{\mu}{s_{n+k}}D^{\mu}{s_{n+k}}+(n+k)(n+k-4){s_{n+k}}{s_{n+k}})
               \nonumber \\
               +&\frac{1}{4}\sum_{p \neq k+2,n-k,n+k}\Big(D_{\mu}{s_{p}}D^{\mu}{s_{p}}+p(p-4){s_{p}}{s_{p}}\Big)             
               \nonumber \\
               +&\frac{1}{2}\sum_p\Big({F_{\mu \nu,p}}^{2}+(p^2-1)(A_{\mu,p})^{2}\Big) 
               \nonumber \\
               +&\frac{1}{4}D_{\mu}\phi_{\nu \rho,p}D^{\mu}\phi_{p}^{\nu \rho}-\frac{1}{2}D_{\mu}\phi_{p}^{\mu \nu}D^{\rho}\phi_{\rho \nu,p}
               +\frac{1}{2}D_{\mu}\phi^{\nu}_{\nu,p}D_{\rho}\phi^{\mu \rho}_{p}
               \nonumber \\
               -&\frac{1}{4}\sum_{p}D_{\mu}\phi^{\nu}_{\nu,p}D^{\mu}\phi^{\rho}_{\rho,p}
               +\frac{(p^2+4p-2)}{4}\phi_{\mu \nu,p}\phi^{\mu \nu}_{p}-\frac{(p^2+4p+2)}{4}(\phi^{\mu}_{\mu,p})^{2}
\end{align}
where $F_{\mu \nu,p}=\partial_{\mu}A_{\nu,p}-\partial_{\nu}A_{\mu p}$, and summation over upper indices is assumed, running over the basis of the irreducible representation corresponding to the field\footnote{We often use the notation $s_p^{I_m}\equiv s_{p}^{m}$.}.  We also haven't specified the values over which $p$ is being summed over. These will be specified by the various selection rules that determines the fields that are exchanged in the bulk. We should finally point out that the fields have been rescaled in order to simplify the action. In this case, the corresponding rescaling factors for scalars and vectors\footnote{The only scalar field that hasn't  been rescaled is $\phi_{p}$.} are given by
\begin{equation}
s_{p} \rightarrow \sqrt{ \frac{(p+1)}{2^{6}p(p-1)(p+2)}} s_{p} \qquad A_{\mu,p} \rightarrow
2\sqrt{\frac{p+2}{p+1}} A_{\mu,p}
\label{normcan}
\end{equation}
and all symmetric tensors are left unscaled. The relevant cubic couplings \cite{Arutyunov:1999en, Lee:1998bxa, Lee:1999pj} are given by
\begin{align}
\mathcal{L}_{3}&= \mathcal{L}_{3}(s_{k+2},s_{n-k}, s_{n+k})
                \nonumber \\  
                 &-   \frac{(2+k)}{8} \sqrt{\frac{(1+k)}{(1+2 k)}}      \langle C_{k+2}^{1}C_{k+2}^{1}C^{3}_{[0,2k,0]}\rangle \Big(D^{\mu}s_{k+2}^{1}D^{\nu}s_{k+2}^{2}\phi^3_{\mu \nu,2k} 
               -\frac{1}{2}(D^{\mu}s_{k+2}^{1}D_{\mu}s_{k+2}^{1}
               \nonumber \\
               &+\left.\frac{1}{2}(2m_{k+2}^2-f(2k)) s_{k+2}^{1}s_{k+2}^{1})\phi^{\nu,3}_{\nu,2k} \right)
               \nonumber \\
                &- \frac{1}{4}  \sqrt{\frac{(1+k)(1+k-n) (k-n) (1+2 k)}{  (-1+k+n) (k+n)}} \langle C_{n-k}^{1}C_{n+k}^{2}C^{3}_{[0,2k,0]}\rangle \left(D^{\mu}s_{n-k}^{1}D^{\nu}s_{n+k}^{2}\phi^3_{\mu \nu,2k}
               -(D^{\mu}s_{n-k}^{1}D_{\mu}s_{n+k}^{1}\right.
               \nonumber \\
               &+\left.\frac{1}{2}(m_{n-k}^2+m_{n+k}^2-f(2k)) s_{n-k}^{1}s_{n+k}^{1})\phi^{\nu,3}_{\nu,2k} \right)
               \nonumber \\
                &-\sqrt{\frac{(1+k) (2+k) (k-n) (1+k-n)}{2^{5}(-1+n) n}} \langle {C_{k+2}}^{1}C_{n-k}^{2}C^{3}_{[0,n-2,0]}\rangle \left(D^{\mu}s_{k+2}^{1}D^{\nu}s_{n-k}^{2}\phi^3_{\mu \nu,n-2}
               -(D^{\mu}s_{k+2}^{1}D_{\mu}s_{n-k}^{2}\right.
               \nonumber \\
               &+\left.\frac{1}{2}(m_{k+2}^2+m_{n-k}^2-f(n-2)) s_{k+2}^{1}s_{n-k}^{2})\phi^{\nu,3}_{\nu,n-2} \right)
               \nonumber \\    
                &-\sqrt{\frac{(1+k) (2+k)(-1+n) n}{ 2^{5}(-1+k+n) (k+n)}} \langle {C_{k+2}}^{1}C_{n+k}^{2}C^{3}_{[0,n-2,0]}\rangle \left(D^{\mu}s_{k+2}^{1}D^{\nu}s_{n+k}^{2}\phi^3_{\mu \nu,n-2}
               -(D^{\mu}s_{k+2}^{1}D_{\mu}s_{n+k}^{2}\right.
               \nonumber \\
               &+\left.\frac{1}{2}(m_{k+2}^2+m_{n+k}^2-f(n-2)) s_{k+2}^{1}s_{n+k}^{2})\phi^{\nu,3}_{\nu,n-2} \right)
               \nonumber \\    
               &- \frac{1}{2} (1+k) (2+k) \langle C^{1}_{k+2}C^{2}_{k+2}C^{3}_{[1,2k,1]}\rangle s_{k+2}^{1}D^{\mu}s_{k+2}^{2}A_{\mu,2k+1}^{3}
                \nonumber \\
               &-\frac{(1+2k)}{2} \sqrt{\frac{(k-n) (1+k-n) (k+n)}{(-1+k+n)}}\langle C^{1}_{n-k}C^{2}_{n+k}C^{3}_{[1,2k,1]}\rangle s_{n-k}^{1}D^{\mu}s_{n+k}^{2}A_{\mu,2k+1}^{3}
               \nonumber \\
               &-\frac{1}{2} \sqrt{(1+k) (2+k) (k-n) (1+k-n)}\langle C^{1}_{k+2}C^{2}_{n-k}C^{3}_{[1,n-2,1]}\rangle s_{k+2}^{1}D^{\mu}s_{n-k}^{2}A_{\mu,n-1}^{3}
               \nonumber \\
               &-\frac{(-1+n)}{2}\sqrt{\frac{(1+k) (2+k)(k+n)}{(-1+k+n)}}\langle C^{1}_{k+2}C^{2}_{n+k}C^{3}_{[1,n-2,1]}\rangle s_{k+2}^{1}D^{\mu}s_{n+k}^{2}A_{\mu,n-1}^{3}
               \nonumber
\end{align}
where 
\begin{equation}
m_{p}^2=p(p-4), \qquad \qquad f(p)=p(p+4)
\end{equation}
As one can see, there are different contributions to the $s$- and $t$-channels. The cubic terms for the fields $s_p^I$ will be explicitly written below given the various symmetry factors that enter depending on the values of $n$ and $k$. We now consider the interaction vertices for the scalars $s_p^I$. One has two different cases. For $k\neq 0$
\begin{align}
&\mathcal{L}_{3}(s_{k+2},s_{n-k},s_{n+k})=
					-\frac{1}{2} (1+k) (2+k) \sqrt{(1+k) (1+2 k)} \langle C^{1}_{k+2}C^{2}_{k+2}C^{3}_{[0,2k+2,0]}\rangle s_{k+2}^{1}s_{k+2}^{2}s_{2k+2}^{3} 
					\nonumber \\
					&- \sqrt{(1+k) (1+2 k) (k-n) (1+k-n) (-1+k+n) (k+n)}\langle C^{1}_{n-k}C^{2}_{n+k}C^{3}_{[0,2k+2,0]}\rangle s_{n-k}^{1}s_{n+k}^{2}s_{2k+2}^{3} 								\nonumber \\
           				&-\sqrt{\frac{(1+k) (2+k) (k-n) (1+k-n) (-1+n) n}{2}}\langle C^{1}_{k+2}C^{2}_{n-k}C^{3}_{[0,n,0]} \rangle s_{k+2}^{1}s_{n-k}^{2}s_{n}^{3} 							\nonumber\\
					&-\sqrt{\frac{(1+k) (2+k) (k+n) (-1+k+n) (-1+n) n}{ 2}}\langle C^{1}_{k+2}C^{2}_{n+k}C^{3}_{[0,n,0]} \rangle s_{k+2}^{1}s_{n+k}^{2}s_{n}^{3} 								\nonumber
\end{align}
and for $k=0$ one gets the case discussed in \cite{Uruchurtu:2008kp}. Finally, the quartic couplings are given by
\begin{equation}
\mathcal{L}_{4}=\mathcal{L}_{4}^{(0)}+\mathcal{L}_{4}^{(2)}+\mathcal{L}_{4}^{(4)}
\label{quarticlag}
\end{equation}
where the supraindex indicates contributions coming from zero, two and four-derivative terms. This is
\begin{equation}
\mathcal{L}_{4}=\mathcal{L}_{k_{1}k_{2}k_{3}k_{4}}^{(0)I_{1}I_{2}I_{3}I_{4}}s_{k_{1}}^{I_{1}}s_{k_{2}}^{I_{2}}s_{k_{3}}^{I_{3}}s_{k_{4}}^{I_{4}}+
\mathcal{L}_{k_{1}k_{2}k_{3}k_{4}}^{(2)I_{1}I_{2}I_{3}I_{4}}s_{k_{1}}^{I_{1}}D_{\mu}s_{k_{2}}^{I_{2}}s_{k_{3}}^{I_{3}}D^{\mu}s_{k_{4}}^{I_{4}}
+\mathcal{L}_{k_{1}k_{2}k_{3}k_{4}}^{(4)I_{1}I_{2}I_{3}I_{4}}s_{k_{1}}^{I_{1}}D_{\mu}s_{k_{2}}^{I_{2}}D^{\nu}D_{\nu}(s_{k_{3}}^{I_{3}}D^{\mu}s_{k_{4}}^{I_{4}}) \nonumber
\end{equation}
The explicit form of these terms has been computed in \cite{Arutyunov:1999fb}. 
For our case, two of the $k_{i}$'s are equal to $k+2$. This allows for twelve possible permutations, where the indices $I_{i}$ run over the basis of the representation $[0,k_{i},0]$ which is being summed over. We show in appendix \ref{sec:QuarticInt} that the relevant contributions coming from the quartic lagrangian can be reduced to a simple expression involving only two-derivative terms and zero-derivative terms, which is consistent with the fact that this is a sub-subextremal process, i.e. $k_1 + k_2 + k_3 - k_4 = 4$, as indicated in \cite{Arutyunov:2000ima, D'Hoker:2000dm}. 
\begin{figure}[ht]
\begin{center}
\resizebox{108mm}{36mm}{\input{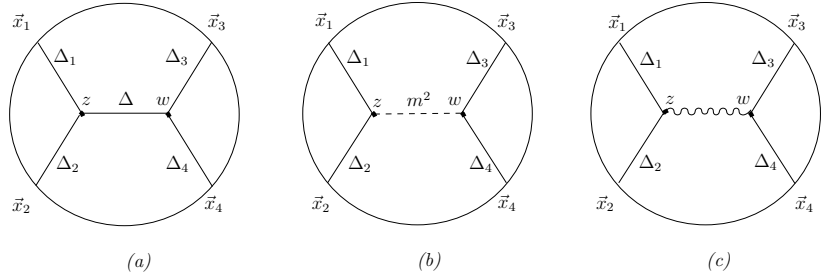}} 
\end{center}
\caption{Witten Diagrams for the $s$-channel process. \emph{(a)} exchange by a scalar with $m^2=(2k+2)(2k-2)$ \emph{(b)} exchange by a massive vector of mass $m_{2k+1}^2=4k(k+1)$  \emph{(c)}  exchange by a tensor field of mass $f(2k)=4k(k+2)$}
\label{schanneldiffw}
\end{figure}
\begin{figure}[ht]
\begin{center}
\resizebox{144mm}{36mm}{\input{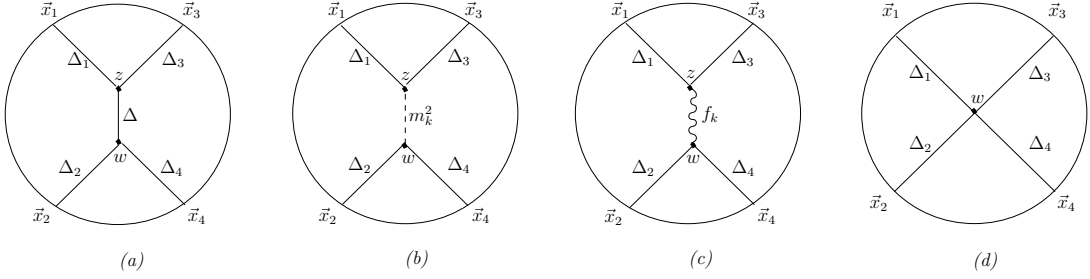}} 
\end{center}
\caption{Witten Diagrams for the $t$-channel process. \emph{(a)} exchange by a scalar of mass $m^2=n(n-4)$ \emph{(b)} exchange by a vector of mass $m_{n-1}^2=n(n-2)$ \emph{(c)} exchange by a tensor field of mass $f(n-2)=(n-2)(n+2)$ \emph{(d)} Contact diagram. \ }
\label{tchanneldiffw}
\end{figure}
Now that the relevant terms in the lagrangian have been specified, it remains to compute its on-shell value. From the couplings, one can determine the diagrams that need to be computed. One has scalar exchanges of $s_{p}^{I}$, vector exchanges $A^{I}_{\mu,[1,p-1,1]}$, and symmetric tensor exchanges, $\phi_{\mu\nu,[0,p-2,0]}$. Furthermore, contact diagrams arise from the quartic couplings discussed above. 

We start by introducing the following currents: 
\begin{align}
J_{\mu,2k+1}^{I}&= \frac{1}{2} (1+k) (2+k) \langle C^{1}_{k+2}C^{2}_{k+2}C^{I}_{[1,2k,1]}\rangle (s_{k+2}^{1}D_{\mu}s_{k+2}^{2}-s_{k+2}^{2}D_{\mu}s_{k+2}^{1}) \nonumber \\
               &+\frac{(1+2k)}{2} \sqrt{\frac{(k-n) (1+k-n) (k+n)}{(-1+k+n)}}\langle C^{1}_{n-k}C^{2}_{n+k}C^{I}_{[1,2k,1]}\rangle (s_{n-k}^{1}D_{\mu}s_{n+k}^{2}-s_{n+k}^{2}D_{\mu}s_{n-k}^{1})
               \nonumber \\
J_{\mu,n-1}^{I}&=\frac{1}{2} \sqrt{(1+k) (2+k) (k-n) (1+k-n)}\langle C^{1}_{k+2}C^{2}_{n-k}C^{I}_{[1,n-2,1]}\rangle (s_{k+2}^{1}D_{\mu}s_{n-k}^{2}-s_{n-k}^{2}D_{\mu}s_{k+2}^{1})
               \nonumber \\
               &+\frac{(n-1)}{2}\sqrt{\frac{(1+k) (2+k)(k+n)}{(-1+k+n)}}\langle C^{1}_{k+2}C^{2}_{n+k}C^{I}_{[1,n-2,1]}\rangle (s_{k+2}^{1}D_{\mu}s_{n+k}^{2}-s_{n+k}^{2}D_{\mu}s_{k+2}^{1})
               \nonumber \\
T^I_{\mu\nu, 2k}&=   \frac{(2+k)}{2} \sqrt{\frac{(1+k)}{(1+2 k)}} \langle C_{k+2}^{1}C_{k+2}^{2}C^{I}_{[0,2k,0]}\rangle \mathcal{T}^{k+2, k+2}_{\mu\nu, 2k}
               \nonumber \\
                &+  \sqrt{\frac{(1+k)(1+k-n) (k-n) (1+2 k)}{  (-1+k+n) (k+n)}} \langle C_{n-k}^{1}C_{n+k}^{2}C^{I}_{[0,2k,0]}\rangle \mathcal{T}^{n-k,n+k}_{\mu\nu, 2k}
               \nonumber \\
T^{I}_{\mu\nu, n-2}&=\sqrt{\frac{(1+k) (2+k) (k-n) (1+k-n)}{2(-1+n) n}} \langle {C_{k+2}}^{1}C_{n-k}^{2}C^{I}_{[0,n-2,0]}\rangle \mathcal{T}^{k+2,n-k}_{\mu\nu, n-2}
               \nonumber \\    
                &+\sqrt{\frac{(1+k) (2+k)(-1+n) n}{ 2(-1+k+n) (k+n)}} \langle {C_{k+2}}^{1}C_{n+k}^{2}C^{I}_{[0,n-2,0]}\rangle \mathcal{T}^{k+2,n+k}_{\mu\nu, n-2}
                \end{align}
where
\begin{align}
\mathcal{T}^{\Delta_1 \Delta_2}_{\mu\nu , p }&= \left(D_{(\mu}s_{\Delta_{1}}^{1}D_{\nu)}s_{\Delta_{2}}^{2}
                   -\frac{1}{2}g_{\mu\nu}(D^{\rho}s_{\Delta_1}^{1}D_{\rho}s_{\Delta_2}^{2}
               +\frac{1}{2}(m_{\Delta_1}^2+m_{\Delta_2}^2-f(p)) s_{\Delta_1}^{1}s_{\Delta_2}^{2}) \right)
\end{align}
Now we can write down the equations of motion for the scalars:
\begin{align}
(D_{\mu}^2-m_{n}^2)s_{n}^{I}=
&-\sqrt{2(k+1) (k+2) (n-k) (n-k-1) (n-1) n}\langle C^{1}_{k+2}C^{2}_{n-k}C^{I}_{[0,n,0]} \rangle s_{k+2}^{1}s_{n-k}^{2}							
\nonumber\\
&-\sqrt{2(k+1) (k+2) (k+n) (n+k-1) (n-1) n}\langle C^{1}_{k+2}C^{2}_{n+k}C^{I}_{[0,n,0]} \rangle s_{k+2}^{1}s_{n+k}^{2} 
\nonumber \\
(D_{\mu}^2-m_{2k+2}^2)s_{2k+2}^{I}=&-(k+1) (k+2) \sqrt{(k+1) (2 k+1)} \langle C^{1}_{k+2}C^{2}_{k+2}C^{I}_{[0,2k+2,0]}\rangle s_{k+2}^{1}s_{k+2}^{2} 			
\nonumber\\
&- 2\sqrt{(k+1) (2 k+1) (n-k) (n-k-1) (n+k-1) (n+k)}\langle C^{1}_{n-k}C^{2}_{n+k}C^{I}_{[0,2k+2,0]}\rangle s_{n-k}^{1}s_{n+k}^{2} 	
\nonumber \\
\end{align}
For the vectors:
\begin{align}
D^{\mu}(D_{\mu}A_{\nu,2k+1}^{I}-D_{\nu}A_{\mu,2k+1}^{I})+4k(k+1)A^{I}_{\nu,2k+1}&=-\frac{1}{4}J_{\nu,2k+1}^{I}
\nonumber \\
D^{\mu}(D_{\mu}A_{\nu,n-1}^{I}-D_{\nu}A_{\mu,n-1}^{I})+n(n-2)A^{I}_{\nu,n-1}&=-\frac{1}{4}J_{\nu,n-1}^{I}
\end{align}
and for the tensors:
\begin{align}
&{W_{\mu\nu}}^{\lambda\rho}[\phi_{\lambda \rho, 2k}]=\frac{1}{4}\left(g_{\mu\rho}g_{\nu\sigma}+g_{\mu\sigma}g_{\nu\rho}-\frac{2}{3}g_{\mu\nu}g_{\rho\sigma}\right)T^{\rho\sigma}_{2k}
\nonumber \\
&{W_{\mu\nu}}^{\lambda\rho}[\phi_{\lambda \rho, n-2}]=\frac{1}{4}\left(g_{\mu\rho}g_{\nu\sigma}+g_{\mu\sigma}g_{\nu\rho}-\frac{2}{3}g_{\mu\nu}g_{\rho\sigma}\right)T^{\rho\sigma}_{n-2}
\end{align}
where ${W_{\mu\nu}}^{\lambda \rho}$ is the modified Ricci operator
\begin{align}
{W_{\mu\nu}}^{\lambda \rho}[\phi_{\lambda\rho,p}]&=-D_{\rho}D^{\rho}\phi_{\mu\nu,p}-D_{\mu}D_{\nu}\phi_{\rho,p}^{\rho}+D_{\mu}D^{\rho}\phi_{\nu\rho,p}+D_{\nu}D^{\rho}\phi_{\mu\rho,p}
\nonumber \\
&-\left[(2-f(p))\phi_{\mu\nu,p}-\frac{6+f(p)}{d-1}g_{\mu\nu}\phi_{\rho,p}^{\rho}\right]
\end{align}
which imply the following equations of motion
\begin{align}
\nabla_\rho \nabla^{\rho} {\phi_{\lambda,2k}}^{\lambda}&=\nabla^{\rho}\nabla^{\lambda}\phi_{\rho\lambda,2k}-\frac{2-f(2k)+(d+1)(2+f(2k))}{1-d}{\phi_{\lambda,2k}}^{\lambda}-\frac{2}{1-d}{T_{\lambda,2k}}^\lambda \nonumber \\
\nabla_\rho \nabla^{\rho} {\phi_{\lambda,n-2}}^{\lambda,I}&=\nabla^{\rho}\nabla^{\lambda}\phi^{I}_{\rho\lambda,n-2}-\frac{2-f(n-2)+(d+1)(2+f(n-2))}{1-d}{\phi_{\lambda,n-2}}^{\lambda,I}-\frac{2}{1-d}{T_{\lambda,n-2}}^\lambda
\end{align}
We now represent the solutions to the equations of motion in the form
\begin{equation}
s_{p}=s_{p}^{0}+\tilde{s}_{p} \qquad A_{\mu}=A_{\mu}^{0}+\tilde{A}_{\mu} \qquad \phi_{\mu\nu}=\phi_{\mu\nu}^{0}+\tilde{\phi}_{\mu\nu}
\end{equation}
where $s^{0}_{p}$, $A_{\mu}^{0}$ and $\phi_{\mu\nu}^{0}$ are solutions to the linearised equations with fixed boundary conditions and $\tilde{s}_{p}$, $\tilde{A}_{\mu}$ and $\tilde{\phi}_{\mu\nu}$ represent the fields in the AdS bulk with vanishing boundary conditions. It is then possible to express these fields in terms of an integral on the bulk, involving the corresponding Green function:
\begin{align}
&(D_\mu^2-m_{p}^2)G_{p}(u)=-\delta(z,w) \nonumber \\
&D^{\rho}(D_{\rho} G_{\mu\nu}(u;p)-D_{\mu}G_{\rho\nu}(u;p))+m_{p}^2G_{\mu\nu}(u;p)=-g_{\mu\nu}\delta(z,w) \nonumber \\
&{W_{\mu\nu}}^{\lambda\omega}[G_{\lambda\omega\rho\sigma}(u;p)]=\left(g_{\mu\rho}g_{\nu\sigma}+g_{\mu\sigma}g_{\nu\rho}-\frac{2}{3}g_{\mu\nu}g_{\rho\sigma}\right)\delta(z,w)
\end{align}
where $u$ is the chordal distance in AdS:
\begin{equation}
u=\frac{(z-w)^2}{2z_0w_0}
\end{equation}
To first order in perturbation theory, the solutions read
\begin{align}
\tilde{s}^{I}_{2k+2}(w)&= 2(k+1) (k+2) \sqrt{(k+1) (2 k+1)} \langle C^{1}_{k+2}C^{2}_{k+2}C^{I}_{[0,2k+2,0]}\rangle \int [dz] G_{2k+2}(z,w) s_{k+2}^{1}(z)s_{k+2}^{2}(z) 
\nonumber \\
&+2\sqrt{(k+1) (2k+1) (n-k)(n+k) (n-k-1) (n+k-1)}\langle C^{3}_{n-k}C^{4}_{n+k}C^{I}_{[0,n,0]} \rangle 
\nonumber \\
& \times \int [dz] G_{2k+2}(z,w) s_{n-k}^{3}(z)s_{n+k}^{4}(z) 				
\nonumber \\
\tilde{s}_{n}^{I}(w)&=\sqrt{2(k+1) (k+2) (n-k) (n-k-1) n(n-1) }\langle C^{1}_{k+2}C^{3}_{n-k}C^{I}_{[0,n,0]} \rangle \int [dz]G_{n}(z,w)s_{k+2}^{1}(z)s_{n-k}^{3}(z) 			
\nonumber\\
&+\sqrt{2(k+1) (k+2) (n+k) (n+k-1) n(n-1)}\langle C^{2}_{k+2}C^{4}_{n+k}C^{I}_{[0,n,0]} \rangle \int [dz] G_{n}(z,w)s_{k+2}^{2}(z)s_{n+k}^{4}(z) 
\nonumber \\
\tilde{A}_{\mu,2k+1}^{I}(w)&=\frac{1}{4} \int [dz] {G_{\mu}}^{\nu}(z,w)J^{I}_{\nu,2k+1}(z) \qquad \qquad \tilde{A}_{\mu,n-1}^{I}(w)=\frac{1}{4} \int [dz] {G_{\mu}}^{\nu}(z,w)J^{I}_{\nu,n-1}(z) 
\nonumber \\
\tilde{\phi}^{I}_{\mu\nu,2k}(w)&=
\frac{1}{4}\int [dz]G_{\mu\nu\mu'\nu'}(z,w)T^{\mu'\nu',I}_{2k}(z)
\qquad \qquad
\tilde{\phi}^{I}_{\mu\nu,n-2}(w)=\frac{1}{4} \int [dz]G_{\mu\nu\mu'\nu'}(z,w)T_{n-2}^{\mu'\nu',I}(z)
\end{align}
and to avoid cluttering, we've omitted additional subindices to denote the appropriate weights of the Green's functions. We will drop the tilde in the following. Using the expressions above, we arrive at the following expression for the on-shell value of the action.  
The contributions to the $s$-channel are determined by
\begin{align}
\mathcal{L}_{\mathrm{s-channel}}=-&\frac{1}{2}(k+1)^{2}(2k+1)(k+2)\sqrt{(n-k)(n+k)(n-k-1)(n+k-1)} \times
\nonumber \\
&\langle C_{k+2}^{1}C_{k+2}^{2}C_{2k+2}^{I}\rangle \langle C_{n-k}^{3}C_{n+k}^{4}C_{2k+2}^{I}\rangle \int [dz][dw] s_{k+2}^{1}(z)s^2_{k+2}(z) G_{2k+2}(z,w)s_{n-k}^{3}(w)s_{n+k}^{4}(w) \nonumber \\
-&\frac{1}{2^{5}}(k+1)(k+2)(2k+1)\sqrt{\frac{(n-k)(n+k)(n-k-1)}{(n+k-1)}}\langle C_{k+2}^{1}C_{k+2}^{2}C_{[1,2k,1]}^{I}\rangle \langle C_{n-k}^{3}C_{n+k}^{4}C_{[1,2k,1]}^{I}\rangle\times
\nonumber \\
&\int [dz][dw] \left\{s_{k+2}^{1}(z)\overleftrightarrow{D}_{\mu}s_{k+2}^{2}(z)G^{\mu\nu}(z,w)s_{n-k}^{3}(w)\overleftrightarrow{D}_{\nu}s_{n+k}^{4}(w)\right\}
 \nonumber \\
&-\frac{1}{2^{5}}(k+2)(k+1)\sqrt{\frac{(n-k)(n-k-1)}{(n+k)(n+k-1)}}\langle C_{k+2}^{1}C_{k+2}^{2}C_{[0,2k,0]}^{I}\rangle \langle C_{n-k}^{3}C_{n+k}^{4}C_{[0,2k,0]}^{I}\rangle\times
\nonumber \\
&\int [dz][dw] \mathcal{T}^{\mu\nu,k+2,k+2}_{2k}(z){G_{\mu\nu}}^{\mu'\nu'}(z,w)\mathcal{T}^{n-k,n+k}_{\mu'\nu',2k}(w)
\label{schannelos}
\end{align}
and for the $t$-channel the analogue expression reads
\begin{align}
\mathcal{L}_{\mathrm{t-channel}}=
-&\frac{1}{2}(k+1)(k+2)n(n-1)\sqrt{(n-k)(n+k)(n-k-1)(n+k-1)} \times 
\nonumber \\
&\langle C_{k+2}^{1}C_{n-k}^{3}C_{n}^{I}\rangle \langle C_{k+2}^{2}C_{n-k}^{4}C_{n}^{I}\rangle \int [dz][dw] s_{k+2}^{1}(z)s_{n-k}^{3}(z) G_{n}(z,w)s_{k+2}^{2}(w)s_{n+k}^{4}(w) \nonumber \\
-&\frac{1}{2^{5}}(n-1)(k+1)(k+2)\sqrt{\frac{(n-k)(n+k)(n-k-1)}{(n+k-1)}}\langle C_{k+2}^{1}C_{n-k}^{3}C_{[1,n-2,1]}^{I}\rangle \langle C_{k+2}^{2}C_{n+k}^{4}C_{[1,n-2,1]}^{I}\rangle\times
\nonumber \\
&\int [dz][dw] \left\{s_{k+2}^{1}(z)\overleftrightarrow{D}_{\mu}s_{n-k}^{3}(z)G^{\mu\nu}(z,w)s_{k+2}^{2}(w)\overleftrightarrow{D}_{\nu}s_{n+k}^{4}(w)\right\}
\nonumber \\
&-\frac{1}{2^{5}}(k+2)(k+1)\sqrt{\frac{(n-k)(n-k-1)}{(n+k)(n+k-1)}}\langle C_{k+2}^{1}C_{n-k}^{3}C_{[0,n-2,0]}^{I}\rangle \langle C_{k+2}^{2}C_{n+k}^{4}C_{[0,n-2,0]}^{I}\rangle\times
\nonumber \\
&\int [dz][dw] \mathcal{T}^{\mu\nu,k+2,n-k}_{n-2}(z){G_{\mu\nu}}^{\mu'\nu'}(z,w)\mathcal{T}^{k+2,n+k}_{\mu'\nu',n-2}(w)
\label{tchannelos}
\end{align}
The expressions in brackets arise from the integrals over $S^5$ and are defined in appendix \ref{sec:integrationsphere}. We will worry about contact interactions later. So far, we see that we need to compute three Witten Diagrams for each channel, involving exchanges of scalars, massless and massive gauge bosons and massless and massive gravitons. In order to do so, we extend the methods developed in \cite{D'Hoker:1999pj,D'Hoker:1999ni,Berdichevsky:2007xd} to perform the computations. We list the explicit results that we will use, which can be derived from (\ref{124eq}) and (\ref{125eq}) in appendix \ref{sec:integrationsphere}.
\begin{align}
\langle C_{k+2}^{1}C_{k+2}^{2}C_{2k+2}^{I}\rangle \langle C_{n-k}^{3}C_{n+k}^{4}C_{2k+2}^{I}\rangle&\rightarrow\frac{1}{2}\sigma+\frac{1}{2}\tau -\frac{k+1}{2(2k+3)} \nonumber \\
\langle C_{k+2}^{1}C_{n-k}^{3}C_{n}^{I}\rangle \langle C_{k+2}^{2}C_{n-k}^{4}C_{n}^{I}\rangle &\rightarrow \frac{(k+1)}{n(n+1)}\left[ -(n-k-1)+(n+1)\sigma+(n+1)(n-k-1)\tau \right] \nonumber  \\
\langle C_{k+2}^{1}C_{k+2}^{2}C_{[1,2k,1]}^{I}\rangle \langle C_{n-k}^{3}C_{n+k}^{4}C_{[1,2k,1]}^{I}\rangle&\rightarrow\frac{2(k+1)}{2k+1}(\sigma-\tau) \nonumber \\
\langle C_{k+2}^{1}C_{n-k}^{3}C_{[1,n-2,1]}^{I}\rangle \langle C_{k+2}^{2}C_{n+k}^{4}C_{[1,n-2,1]}^{I}\rangle&\rightarrow\frac{n}{(n+2)(n-1)}\left[-(n-2k-2)+(n+2)(\sigma-\tau)\right] \nonumber \\
\langle C_{k+2}^{1}C_{k+2}^{2}C_{[0,2k,0]}^{I}\rangle \langle C_{n-k}^{3}C_{n+k}^{4}C_{[0,2k,0]}^{I}\rangle &\rightarrow1\nonumber \\
\langle C_{k+2}^{1}C_{n-k}^{3}C_{[0,n-2,0]}^{I}\rangle \langle C_{k+2}^{2}C_{n+k}^{4}C_{[0,n-2,0]}^{I}\rangle&\rightarrow1
\end{align}
\subsection{Results for Exchange Integrals}
\begin{figure}[t]
\begin{center}
\resizebox{60mm}{38mm}{\input{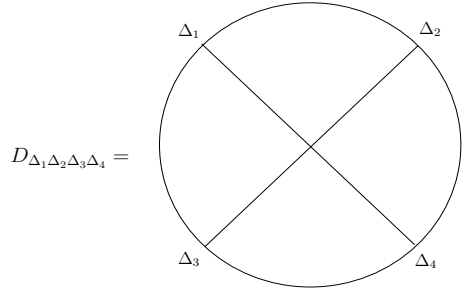}} 
\end{center}
\caption{Graphic representation of a $D$-function.}
\label{wittendfunc}
\end{figure}

We now carry out the integrals and write the results in terms of $\bar{D}$-functions, which are functions of $u$ and $v$ introduced in (\ref{crossradii}) and are related to the more familiar $D$-functions \cite{D'Hoker:1999pj} which are defined as
\begin{equation}
D_{\Delta_1\Delta_2\Delta_3,\Delta_4}(\vec{x}_1,\vec{x}_2,\vec{x}_3,\vec{x}_4)=\int [dw] \tilde{K}_{\Delta_1}(w,\vec{x}_1)\tilde{K}_{\Delta_2}(w,\vec{x}_2)\tilde{K}_{\Delta_3}(w,\vec{x}_3)\tilde{K}_{\Delta_4}(w,\vec{x}_4) 
\end{equation}
where $\tilde{K}_{\Delta}(w,\vec{x})$ is the normalised bulk-to-boundary propagator for a scalar of conformal dimension $\Delta$
\begin{equation}
\tilde{K}_{\Delta}(z,\vec{x})=\left(\frac{z_{0}}{z_{0}^2+(\vec{z}-\vec{x})^2}\right)^{\Delta}
\end{equation}
$D_{\Delta_1\Delta_2\Delta_3\Delta_4}$ can be identified as a quartic scalar interactions (see Fig. \ref{wittendfunc}). The relation between the $D$-functions and the $\bar{D}$-functions, and their properties can be found in appendix \ref{sec:Dfunc}. 

Let us first introduce the following notation for the various exchange integrals that contribute to the amplitude. 
\begin{align}
S_{\Dl_{1}\Dl_{2}\Dl_{3}\Dl_{4}}(\vec{x}_{1},\vec{x}_{2},\vec{x}_{3},\vec{x}_{4})&=\int [dw] [dz]\tilde{K}_{\Dl_{1}}(z,\vec{x}_{1})\tilde{K}_{\Dl_{2}}(z,\vec{x}_{2})G(z,w)\tilde{K}_{\Dl_{3}}(w,\vec{x}_{3})\tilde{K}_{\Dl_{4}}(w,\vec{x}_{4})
\nonumber \\
V_{\Dl_{1}\Dl_{2}\Dl_{3}\Dl_{4}}(\vec{x}_{1},\vec{x}_{2},\vec{x}_{3},\vec{x}_{4})&=\int [dw] [dz]\tilde{K}_{\Dl_{1}}(z,\vec{x}_{1})\buildrel\leftrightarrow\over{D^{\mu}}\tilde{K}_{\Dl_{2}}z,\vec{x}_{2})
G_{\mu\nu}(z,w)\tilde{K}_{\Dl_{3}}(w,\vec{x}_{3})\buildrel\leftrightarrow\over{D^{\nu}}\tilde{K}_{\Dl_{4}}(w,\vec{x}_{4}) 
\nonumber \\
T_{\Dl_{1}\Dl_{2}\Dl_{3}\Dl_{4}}(\vec{x}_{1},\vec{x}_{2},\vec{x}_{3},\vec{x}_{4})&=\int [dz] [dw] T^{\mu\nu}_{\Dl_{1}\Dl_{2}}(z,\vec{x}_{1},\vec{x}_{2})G_{\mu\nu\mu'\nu'}(z,w)T^{\mu'\nu'}_{\Dl_{3},\Dl_{4}}(w,\vec{x}_{3},\vec{x}_{4}) 
\end{align}
with the bulk-to-bulk propagators appropriately chosen, depending on the particle that is being exchanged.  For our case, the $s$-channel integrals yield
\begin{align}
S_{k+2\ k+2\ n-k\ n+k}=&\frac{\pi^2}{8}\frac{1}{(n+k-1)\Gamma(k+2)^2\Gamma(n-k)} 
\frac{u\bar{D}_{k+1\ k+1\ n-k \ n+k}}{{|\vec{x}_{14}|^{2k}|\vec{x}_{24}|^{2k}|\vec{x}_{12}|}^{4}{|\vec{x}_{34}|}^{2(n-k)}}
\nonumber \\
V_{k+2\ k+2\ n-k\ n+k}=&-\frac{\pi^2}{4\Gamma(k+2)^2\Gamma(n-k)}\frac{u}{{|\vec{x}_{14}|^{2k}|\vec{x}_{24}|^{2k}|\vec{x}_{12}|}^{4}{|\vec{x}_{34}|}^{2(n-k)}}
\left\{\bar{D}_{k+1\ k+2\ n-k\ n+k+1}\right.
\nonumber \\
&+\left.\bar{D}_{k+2\ k+1\ n-k+1\ n+k}-v\bar{D}_{k+1\ k+2\ n-k+1\ n+k}-\bar{D}_{k+2 \ k+1\ n-k\ n+k+1}\right\}
\nonumber \\
T_{k+2\ k+2\ n-k\ n+k}=&-\frac{\pi^2(n+k)}{(2k+3)\Gamma(k+2)^2\Gamma(n-k)}\frac{u}{{|\vec{x}_{14}|^{2k}|\vec{x}_{24}|^{2k}|\vec{x}_{12}|}^{4}{|\vec{x}_{34}|}^{2(n-k)}}
\left\{(k+1)\left(v\bar{D}_{k+1\ k+3\ n-k+1\ n+k+1}\right.\right. 
\nonumber \\
&+\left.\bar{D}_{k+3 \ k+1\ n-k+1\ n+k+1}\right)-(k+2)(n-k-2)u\bar{D}_{k+2\ k+2\ n-k\ n+k} 
\nonumber \\
&-\left.(k+2)(1+v-u)\bar{D}_{k+2\ k+2\ n-k+1\ n+k+1}\right\} 
\nonumber \\
\label{schannresults}
\end{align}
and the $t$-channel amplitudes are given by
\begin{align}
S_{k+2\ n-k\ k+2\ n+k}=&\frac{\pi^2}{8}\frac{1}{\Gamma(k+2)^2(n+k-1)\Gamma(n-k)}\frac{u^2\bar{D}_{k+1\ k+2\ n-k-1\ n+k}}{{|\vec{x}_{14}|^{2k}|\vec{x}_{24}|^{2k}|\vec{x}_{12}|}^{4}{|\vec{x}_{34}|}^{2(n-k)}}
\nonumber \\
V_{k+2\ n-k\ k+2\ n+k}=&-\frac{\pi^2}{2n\Gamma(k+2)^2\Gamma(n-k)}\frac{u^2}{{|\vec{x}_{14}|^{2k}|\vec{x}_{24}|^{2k}|\vec{x}_{12}|}^{4}{|\vec{x}_{34}|}^{2n}}
\left\{-n\bar{D}_{k+2\ k+2\ n-k-1\ n+k+1}\right.
\nonumber \\
&+\left.(k+1)\bar{D}_{k+1\ k+2\ n-k\ n+k+1}+(n-k-1)u\bar{D}_{k+2\ k+3\ n-k-1\ n+k}+k(k+1)\bar{D}_{k+1\ k+2\ n-k-1\ n+k}\right\}
\nonumber \\
T_{k+2\ n-k\ k+2\ n+k}=&-\frac{\pi^2}{\Gamma(k+2)^2\Gamma(n-k)}\left[\frac{2(n+k)(n-k)}{(n+1)(n+2)}\right]\frac{u^2}{{|\vec{x}_{14}|^{2k}|\vec{x}_{24}|^{2k}|\vec{x}_{12}|}^{4}{|\vec{x}_{34}|}^{2(n-k)}}
\left\{\frac{(k+1)(k+2)}{(n-k)}\right. \times
\nonumber \\
&v\bar{D}_{k+1\ k+3\ n-k+1\ n+k+1}+(n-k-1)u\bar{D}_{k+3\ k+3\  n-k-1\ n+k+1}
\nonumber \\
&-\left.(k+2)(u+v-1)\bar{D}_{k+2\ k+3\ n-k\ n+k+1}-k(k+2)\bar{D}_{k+2\ k+2\ n-k\ n+k}\right\}
\label{rchannresults}
\end{align}
These expressions are to be substituted in the action, including an overall constant normalisation factor of $C(n-k)C(n+k)C(k+2)^2$ which arises from the bulk-to-boundary propagators $K_{\Delta}(z,\vec{x})$. Here
\begin{equation}
C(n)=\begin{cases}
              \frac{\Gamma(n)}{\pi^{2}\Gamma(n-2)}, \qquad n> 2 \\
              \frac{1}{\pi^2}, \hspace{16mm} n=2
\end{cases}
\end{equation}
\subsection{Contact Interactions}
In appendix \ref{sec:QuarticInt} we explicitly show that the four-derivative terms in the quartic langrangian (\ref{quarticlag}) can be expressed in terms of two- and zero-derivative terms, so the lagrangian is of the $\sigma$-model type. This was expected given that the relevant couplings are next-next-to-extremal \cite{Arutyunov:2000ima}. 

Let us then start from the recast version of the quartic lagrangian: 
\begin{align}
\mathcal{L}_{4}=c_{q}&\Big[A_{2}^{1234}(s_{k+2}^{1}\nabla_{\mu}s_{k+2}^{2}s_{n-k}^{3}\nabla^{\mu}s_{n+k}^{4}+s_{k+2}^{1}\nabla_{\mu}s_{k+2}^{2}s_{n+k}^{3}\nabla^{\mu}s_{n-k}^{4}) \nonumber \\
+&A_{0}^{1234}s_{k+2}^{1}s_{k+2}^{2}s_{n-k}^{3}s_{n+k}^{4}\Big] \nonumber \\
\label{quarticlagfinal}
\end{align}
where
\begin{equation}
c_{q}=\frac{1}{16}(k+1)(k+2)\sqrt{\frac{(n-k)(n-k-1)}{(n+k)(n+k-1)}}
\end{equation}
and
\begin{align}
A_{2}&= -\tau+\sigma\tau \nonumber \\
A_{0}&=-(k+1)(n+k)+(4k^{2}+k(2+7n)+n(n+6))\sigma+n(3n+k-4)\sigma\tau - n(n+k)\sigma^{2}
\end{align}

We evaluate the on-shell value of the lagrangian and use the following identity:
\begin{equation}
D_{\mu}\tilde{K}_{\Delta_{1}}(z,\vec{x}_{1})D^{\mu}\tilde{K}_{\Delta_{2}}(z,\vec{x}_{2})=\Delta_{1}\Delta_{2}\left(\tilde{K}(z,\vec{x}_{1})\tilde{K}(z,\vec{x}_{2})-2|\vec{x}_{12}|^{2}\tilde{K}(z,\vec{x}_{1})\tilde{K}(z,\vec{x}_{1})\right)
\end{equation}
 Rewritting (\ref{quarticlagfinal}) in terms of $\bar{D}$-functions, one obtains
\begin{align}
\mathcal{L}_{4}^{\mathrm{on-shell}}&=\frac{\pi^{2}c_{q}C(k+2)^{2}C(n-k)C(n+k)}{2\Gamma(k+2)^{2}\Gamma(n-k)}\frac{u^{2}(n+k)}{|\vec{x}_{14}|^{2k}|\vec{x}_{24}|^{2k}|\vec{x}_{34}|^{2(n-k)}|\vec{x}_{12}|^{4}}
\nonumber \\
&\times\Big[(-\tau+\sigma \tau)\Big((k+2)\bar{D}_{k+2\  k+2 \ n-k \ n+k}-2\bar{D}_{k+2 \ k+3 \ n-k \ n+k+1}\Big)
\nonumber \\
&+(-\sigma+\sigma\tau)\left((k+2)\frac{(n-k)}{(n+k)}\bar{D}_{k+2\ k+2\ n-k\ n+k}-2v\bar{D}_{k+2\ k+3\ n-k+1 \ n+k}\right)\nonumber \\
&+A_{0}\frac{\bar{D}_{k+2\ k+2\ n-k \ n+k}}{k+2}\Big]
\label{quarticons}
\end{align}
so the final result for the on-shell action will be given by adding the contributions from equations (\ref{schannelos}), (\ref{tchannelos}) and (\ref{quarticons}).
\subsection{Results for the Four Point Function}
We now evaluate the action on-shell by substituting the summation of overlapping $SO(6)$ tensors in terms of $\sigma$ and $\tau$ and the results for the exchange integrals.
\begin{align}
\mathcal{S}&=\frac{N^{2}}{8\pi^{8}}\frac{\Gamma(n+k)}{\Gamma(k)^{2}\Gamma(n-k-2)\Gamma(n+k-2)}
\int d^{4}\vec{x}_{1}d^{4}\vec{x}_{2}d^{4}\vec{x}_{3}d^{4}\vec{x}_{4}
\frac{s_{k+2}(\vec{x}_{1})s_{k+2}(\vec{x}_{2})s_{n-k}(\vec{x}_{3})s_{n+k}(\vec{x}_{4})}{|\vec{x}_{14}|^{2k}|\vec{x}_{24}|^{2k}|\vec{x}_{34}|^{2(n-k)}|\vec{x}_{12}|^{4}}
u\Big\{
\nonumber \\
&+ \frac{1 }{2^{5}}(k+1)(k+2)\sqrt{\frac{(n-k)(n+k)(n-k-1)}{(n+k-1)}}\left[\frac{(k+1)^{2}(2k+1)}{2k+3}\bar{D}_{k+1k+1n-kn+k}\right.
\nonumber \\
&+\frac{1}{(2k+3)}\left( (k+1)\left( v\bar{D}_{k+1k+3n-k+1n+k+1}+\bar{D}_{k+3k+1n-k+1n+k+1} \right) -(k+2)(1+v-u)\bar{D}_{k+2k+2n-k+1n+k+1}\right)
\nonumber \\
&-\left.\left(\frac{(k+2)(n-k-2)}{(2k+3)}+(k+1)\right)u\bar{D}_{k+2k+2n-kn+k}\right]
\nonumber \\
&+\frac{\sigma}{2^{5}}(k+1)(k+2)\sqrt{\frac{(n-k)(n+k)(n-k-1)}{(n+k-1)}}\left[(k+1) \left( -2(2k+1)\bar{D}_{k+1k+1n-kn+k}+\bar{D}_{k+1k+2n-kn+k+1}\right.\right.
\nonumber \\
&+\left.\bar{D}_{k+2k+1n-k+1n+k}-v\bar{D}_{k+1k+2n-k+1n+k}-\bar{D}_{k+2k+1n-kn+k+1}-2(n-1)u\bar{D}_{k+1k+2n-k-1n+k}\right)
\nonumber \\
&+u \left( -n\bar{D}_{k+2k+2n-k-1n+k+1}+(k+1)\bar{D}_{k+1k+2n-kn+k+1}+(n-k-1)u\bar{D}_{k+2k+3n-k-1n+k} \right.
\nonumber \\
&+\left. k(k+1)\bar{D}_{k+1k+2n-k-1n+k}\right) +\left( 4k^{2}+k(2+7n)+n(n+6)+(n-k)(k+2) \right) u\frac{\bar{D}_{k+2k+2n-kn+k}}{n+k}
\nonumber \\
&-\left.2uv \bar{D}_{k+2k+3n-k+1n+k}\right]
\nonumber \\
&-\frac{\tau}{2^{5}}(k+1)(k+2)\sqrt{\frac{(n-k)(n+k)(n-k-1)}{(n+k-1)}}u \left[(k+2)\bar{D}_{k+2k+2n-kn+k}-2\bar{D}_{k+2k+3n-kn+k+1}\right]
\nonumber  \\
&+\frac{\sigma^{2}}{2^5}(k+1)(k+2)\sqrt{\frac{(n-k)(n+k)(n-k-1)}{(n+k-1)}}\left[\frac{2(n-k-1)(n-1)(k+1)}{(n+1)}u\bar{D}_{k+1k+2n-k-1n+k} \right.
\nonumber 
\end{align}
\begin{align}
&+\frac{(2+2k-n)}{(n+2)}u\left[ -n\bar{D}_{k+2k+2n-k-1n+k+1}+(k+1)\bar{D}_{k+1k+2n-kn+k+1}+(n-k-1)u\bar{D}_{k+2k+3n-k-1n+k} \right.
\nonumber \\
&+\left. k(k+1)\bar{D}_{k+1k+2n-k-1n+k}\right]-n u\bar{D}_{k+2k+2n-kn+k} + \frac{2(n-k)u}{(n+1)(n+2)}\left[ \frac{(k+1)(k+2)}{(n-k)}v\bar{D}_{k+1k+3n-k+1n+k+1}\right.
\nonumber \\
&+\left.\left.(n-k-1)u\bar{D}_{k+3k+3n-k-1n+k+1}-(k+2)(u+v-1)\bar{D}_{k+2k+3n-kn+k+1}-k(k+2)\bar{D}_{k+2k+2n-kn+k}\right]\right]
\nonumber \\
&+\frac{\sigma\tau}{2^{5}}(k+1)(k+2)\sqrt{\frac{(n-k)(n+k)(n-k-1)}{(n+k-1)}}\left[-2(n-k-1)(n-1)(k+1)u\bar{D}_{k+1k+2n-k-1n+k}\right.
\nonumber \\
&-u\left( -n\bar{D}_{k+2k+2n-k-1n+k+1}+(k+1)\bar{D}_{k+1k+2n-kn+k+1}+(n-k-1)u\bar{D}_{k+2k+3n-k-1n+k} \right.
\nonumber \\
&+\left. k(k+1)\bar{D}_{k+1k+2n-k-1n+k}\right)-2u \left(\bar{D}_{k+2k+3n-kn+k+1}+v\bar{D}_{k+2k+3n-k+1n+k}\right)
\nonumber \\
&+\left.\left( 2n(k+2)+n(3n+k-4)\right)\right]u\frac{\bar{D}_{k+2k+2n-kn+k}}{n+k}\Big\}
\end{align}
Here we have also used the symmetry $1 \leftrightarrow 2$. It is implied in this expression that the scalar fields refer to the boundary sources, so they depend on the $\vec{x}_i$ coordinates. We are now ready to compute the four point function (\ref{4pfunction2do}) using the AdS/CFT prescription given in (\ref{prescriptionadscft}). To recover the canonical normalisation, let us briefly consider the two-point functions of the operators in the process. Following \cite{Freedman:1998tz}, for a supergravity action of the form
\begin{equation}
S=\frac{N^2}{8\pi^2} \int [dz] \sqrt{-g}\frac{1}{4} \left[  D_{\mu}s_p D^{\mu}s_p+m^2 s_p s_p \right]
\end{equation}
for $p>2$ the corresponding two-point function is given by
\begin{equation}
\langle \mathcal{O}_{p}(\vec{x}_1) \mathcal{O}_{p}(\vec{x}_2)\rangle = \frac{N^2}{(2\pi)^4} \frac{(p-2)\Gamma(p)}{\Gamma(p-2)}\frac{1}{\vert \vec{x}_{12}\vert^{2p}}
\end{equation}
so we re-scale the operators as
\begin{equation}
O_{p}(\vec{x})=\frac{(2\pi)^2}{N} \sqrt{\frac{\Gamma(p-2)}{(p-2)\Gamma(p)}}\mathcal{O}_p(\vec{x})
\end{equation}
This implies that the four point function is of order $\mathcal{O}(1/N^2)$. The explicit form can be determined from
\begin{align}
\langle O_{k+2}(\vec{x}_{1})O_{k+2}(\vec{x}_{2})O_{n-k}(\vec{x}_{3})O(\vec{x}_{4})\rangle&=
\frac{(2\pi)^{8}}{2N^4}\sqrt{\frac{\Gamma(k)^2\Gamma(n-k-2)\Gamma(n+k-2)}{k^2(n-k-2)(n+k-2)\Gamma(k+2)^2\Gamma(n-k)\Gamma(n+k)}}  \nonumber \\
&\times \frac{\delta}{\delta s_{k+2}(\vec{x}_{1})}\frac{\delta}{\delta s_{k+2}(\vec{x}_{2})}
\frac{\delta}{\delta s_{n-k}(\vec{x}_{3})}\frac{\delta}{\delta s_{n+k}(\vec{x}_{4})}(-S)
\end{align}
Upon functional differentiation, and using the symmetries underlying $\sigma$ and $\tau$, we obtain can write the final result as in (\ref{structure4pCten})
with the functions $(a, b_{1}, b_{2}, c_{1}, c_{2}, d)$ given by 
\begin{equation}
(a, b_{1}, b_{2}, c_{1}, c_{2}, d)= -\frac{\sqrt{(k+2)^2(n-k)(n+k)}}{2N^2\Gamma(k+1)^2\Gamma(n-k-1)}(\tilde{a},\tilde{b}_{1},\tilde{b}_{2},\tilde{c}_1,\tilde{c}_2, \tilde{d})
\label{resultdiffweightsugra}
\end{equation}
and
\begin{align}
\tilde{a}(u,v)=&
\frac{2}{(2k+3)}(k+1)^{2}(2k+1)u\bar{D}_{k+1\ k+1\ n-k\ n+k}
\nonumber \\
&+\frac{u}{(2k+3)}\left( (k+1)\left( v\bar{D}_{k+1\ k+3\ n-k+1\ n+k+1}+\bar{D}_{k+3\ k+1\ n-k+1\ n+k+1} \right) \right.
\nonumber \\
&-\left.(k+2)(1+v-u)\bar{D}_{k+2\ k+2\ n-k+1\ n+k+1}\right)
\nonumber \\
&-\left(\frac{(k+2)(n-k-2)}{(2k+3)}+(k+1)\right)2u^{2}\bar{D}_{k+2\ k+2\ n-k\ n+k}
\label{along}
\end{align}
\begin{align}
\tilde{b}_1(u,v)=& -2(k+1)(2k+1)u\bar{D}_{k+1\ k+1\ n-k\ n+k}-2(n-1)(k+1)u^{2}\bar{D}_{k+1\ k+2 \ n-k-1 \ n+k} 
\nonumber \\
&+(k+1)u\left( \bar{D}_{k+1 \ k+2 \ n-k \ n+k+1} +\bar{D}_{k+2 \ k+1 \ n-k+1 \ n+k}-v\bar{D}_{k+1 \ k+2 \ n-k+1 \ n+k} \right.
\nonumber \\
&-\left. \bar{D}_{k+2 \ k+1 \ n-k \ n+k+1}\right) 
\nonumber \\
&+u^{2}\left( -n \bar{D}_{k+2 \ k+2 \ n-k-1 \ n+k+1 }+(k+1)\bar{D}_{k+1 \ k+2 \ n-k \ n+k+1} +(n-k-1)u\bar{D}_{k+2 \ k+3 \ n-k-1 \ n+k} \right.
\nonumber \\
&+\left. k(k+1)\bar{D}_{k+1 \ k+2 \ n-k-1 \ n+k} \right)+(n+2k+2)u^{2}\bar{D}_{k+2 \ k+2 \ n-k \ n+k} 
\nonumber \\
&+4u^{2} \bar{D}_{k+3 \ k+2 \ n-k \ n+k+1} 
\label{b1long}
\end{align}
\begin{align}
\tilde{b}_2(u,v)=& -2(k+1)(2k+1)u\bar{D}_{k+1\ k+1\ n-k\ n+k}-2(n-1)(k+1)u^{2}v^{-1}\bar{D}_{k+2\ k+1 \ n-k-1 \ n+k} 
\nonumber \\
&+(k+1)u\left( \bar{D}_{k+2 \ k+1 \ n-k \ n+k+1} +v^{-1}\bar{D}_{k+1 \ k+2 \ n-k+1 \ n+k}-\bar{D}_{k+2 \ k+1 \ n-k+1 \ n+k} \right.
\nonumber \\
&-\left. \bar{D}_{k+1 \ k+2 \ n-k \ n+k+1}\right) 
\nonumber \\
&+u^{2}v^{-1}\left( -n \bar{D}_{k+2 \ k+2 \ n-k-1 \ n+k+1 }+(k+1)\bar{D}_{k+2 \ k+1 \ n-k \ n+k+1} +(n-k-1)u\bar{D}_{k+3 \ k+2 \ n-k-1 \ n+k} \right.
\nonumber \\
&+\left. k(k+1)\bar{D}_{k+2 \ k+1 \ n-k-1 \ n+k} \right)+(n+2k+2)u^{2}\bar{D}_{k+2 \ k+2 \ n-k \ n+k} 
\nonumber \\
&+4u^{2} \bar{D}_{k+2 \ k+3 \ n-k \ n+k+1} 
\label{b2long}
\end{align}
\begin{align}
\tilde{c}_1(u,v)=&
\frac{2(n-k-1)(n-1)(k+1)}{(n+1)}u^{2}\bar{D}_{k+1\ k+2\ n-k-1\ n+k} +\frac{(2+2k-n)}{(n+2)}k(k+1)u^{2}\bar{D}_{k+1\ k+2\ n-k-1\ n+k} 
\nonumber \\
&+\frac{(2+2k-n)}{(n+2)}u^{2}\left( -n\bar{D}_{k+2\ k+2\ n-k-1\ n+k+1}+(k+1)\bar{D}_{k+1\ k+2 \ n-k\ n+k+1} \right.
\nonumber \\
&+\left. (n-k-1)u\bar{D}_{k+2 \ k+3\ n-k-1\ n+k}\right) -nu^{2}\bar{D}_{k+2 \ k+2 \ n-k \ n+k}
\nonumber \\
&-\frac{2(n-k)u^{2}}{(n+1)(n+2)}\left( \frac{(k+1)(k+2)}{(n-k)}v\bar{D}_{k+1\ k+3\ n-k+1\ n+k+1}\right.
\nonumber \\
&+(n-k-1)u\bar{D}_{k+3\ k+3\ n-k-1\ n+k+1}-(k+2)(u+v-1)\bar{D}_{k+2\ k+3\ n-k\ n+k+1}
\nonumber \\
&- k(k+2)\bar{D}_{k+2\ k+2\ n-k\ n+k}\Big) 
\label{c1long}
\end{align}
\begin{align}
\tilde{c}_2(u,v)=&
\frac{2(n-k-1)(n-1)(k+1)}{(n+1)}u^{2}v^{-1}\bar{D}_{k+2\ k+1\ n-k-1\ n+k} +\frac{(2+2k-n)}{(n+2)}k(k+1)u^{2}v^{-1}\bar{D}_{k+2\ k+1\ n-k-1\ n+k} 
\nonumber \\
&+\frac{(2+2k-n)}{(n+2)}u^{2}v^{-1}\left( -n\bar{D}_{k+2\ k+2\ n-k-1\ n+k+1}+(k+1)\bar{D}_{k+2\ k+1 \ n-k\ n+k+1} \right.
\nonumber \\
&+\left. (n-k-1)u\bar{D}_{k+3 \ k+2\ n-k-1\ n+k}\right) -nu^{2}\bar{D}_{k+2 \ k+2 \ n-k \ n+k}
\nonumber \\
&-\frac{2(n-k)u^{2}}{(n+1)(n+2)}\left( \frac{(k+1)(k+2)}{(n-k)}v^{-1}\bar{D}_{k+3\ k+1\ n-k+1\ n+k+1}\right.
\nonumber \\
&+(n-k-1)uv^{-1}\bar{D}_{k+3\ k+3\ n-k-1\ n+k+1}-(k+2)(u-v+1)v^{-1}\bar{D}_{k+3\ k+2\ n-k\ n+k+1}
\nonumber \\
&- k(k+2)\bar{D}_{k+2\ k+2\ n-k\ n+k}\Big)  
\label{c2long}
\end{align}
\begin{align}
\tilde{d}(u,v)=&-2(n-k-1)(n-1)(k+1)\frac{u^{2}}{v}\left(v\bar{D}_{k+1\ k+2\ n-k-1\ n+k}+\bar{D}_{k+2\ k+1\ n-k-1\ n+k}\right)
\nonumber \\
&+\frac{u^{2}}{v}\Big[ n\left(v\bar{D}_{k+2\ k+2\ n-k-1\ n+k+1}+\bar{D}_{k+2 \ k+2\ n-k-1\ n+k+1}\right)
\nonumber \\
&-(k+1)\left(v\bar{D}_{k+1\ k+2\ n-k\ n+k+1}+\bar{D}_{k+2\ k+1\ n-k\ n+k+1}\right)
\nonumber \\
&-(n-k-1)u\left(\bar{D}_{k+3\ k+2\ n-k-1\ n+k}+v\bar{D}_{k+2\ k+3\ n-k-1\ n+k} \right) \Big]
\nonumber \\
&-\frac{2u^{2}}{v} \Big[ v\left(\bar{D}_{k+2\ k+3\ n-k\ n+k+1}+\bar{D}_{k+3\ k+2\ n-k\ n+k+1}\right)
\nonumber \\
&+v\left(v\bar{D}_{k+2\ k+3 \ n-k+1\ n+k}+\bar{D}_{k+3\ k+2\ n-k+1\ n+k}\right)\Big]+6nu^{2}\bar{D}_{k+2\ k+2\ n-k\ n+k}
\label{dlong}
\end{align}
Using the expressions in (\ref{resultdiffweightsugra}) it is possible to see that the crossing symmetries are respected and that the overall form of the four point amplitude is consistent with conformal symmetry. 
\section{Verifying the CFT predictions}
\label{sec:CFTpredictions}
At a glance, the supergravity result might seem tedious to look at, but the lengthy expressions in (\ref{along})-(\ref{dlong}) can be simplified as shown in appendix \ref{sec:Dfuncsimp}. The simplified coefficient functions are in fact given by
\begin{align}
\tilde{a}(u,v)&=2u \bar{D}_{k+2 \ k+2 \ n-k \ n+k+2} \nonumber \\
\tilde{b}_{1}(u,v)&=-2u \Gamma(k+1)\Gamma(k+2)\Gamma(n-k-1)+2u(v-u-1)\bar{D}_{k+2 \ k+2 \ n-k \ n+k+2} \nonumber \\
\tilde{b}_{2}(u,v)&=-\frac{2u}{v} \Gamma(k+1)\Gamma(k+2)\Gamma(n-k-1)-\frac{2u}{v}(v+u-1)\bar{D}_{k+2 \ k+2 \ n-k \ n+k+2} \nonumber \\
\tilde{c}_{1}(u,v)&=2u^{2}\bar{D}_{k+2 \ k+2 \ n-k \ n+k+2} \qquad \qquad \tilde{c}_{2}(u,v)=2u^{2}v^{-1}\bar{D}_{k+2 \ k+2 \ n-k \ n+k+2} \nonumber \\
\tilde{d}(u,v)&=-\frac{2u^{2}}{v}\Gamma(k+1)^2\Gamma(n-k)+\frac{2u^{2}}{v}(u-v-1)\bar{D}_{k+2 \ k+2 \ n-k \ n+k+2}
\label{allshort}
\end{align}
As we saw in section \ref{sec:FreeFieldTh}, the symmetries of the theory demand the four point correlation function to acquire the form
(\ref{structure4pCten}). We can read out the strongly coupled limit of $\mathcal{G}(u,v;\sigma,\tau)$ by rewriting (\ref{resultdiffweightsugra}) using the simplified expressions in (\ref{allshort})
\begin{align}
\mathcal{G}(u,v;\sigma,\tau)&=\frac{\sqrt{(k+2)^2(n-k)(n+k)}}{N^2}\left\{ (k+1)\left(\sigma u +\tau \frac{u}{v}\right)+(n-k-1)\sigma\tau \frac{u^2}{v} \right. \nonumber \\
&-\left.\frac{1}{\Gamma(k+1)^2\Gamma(n-k-1)}s(u,v;\sigma, \tau) u^{n-k} v^k\bar{D}_{n-k\ n+k+2\ k+2\ k+2}(u,v)\right\}
\label{ampfilnal}
\end{align}
which as expected satisfies (\ref{crosssym4}). Notice that according to the splitting in (\ref{fullamp}), the ``free'' piece $\mathcal{G}_{0}(u,v;\sigma,\tau)$ computed in supergravity partially agrees with the free field theory results in (\ref{freeresultamp}). We can immediately read out the dynamical piece $\mathcal{H}_{I}(u,v)$
\begin{equation}
\mathcal{H}_{I}(u,v)=\frac{\sqrt{(k+2)^2(n-k)(n+k)}}{N^2\Gamma(k+1)^2\Gamma(n-k-1)}u^{n-k}v^{k}\bar{D}_{n-k \ n+k +2 \ k+2 \ k+2}(u,v)
\label{dynamicalschannel}
\end{equation}
We see then, that the full four point correlation function depends solely on an term reminiscent of an effective quartic interaction vertex, namely a unique $\bar{D}$-function. This result supports the conjecture made in \cite{Dolan:2006ec} and supports the field-theoretic arguments discussed in \cite{Arutyunov:2002fh} for the partially non-renormalised form of four point amplitudes of CPOs.
\section{Discussion}
\label{sec:discuss}

In general, a scalar quartic diagram can be decomposed in a regular
part and a singular part \cite{Dolan:2000ut} where each term is given by series expansions in powers of $u$ and
$1-v$. Namely,
\begin{equation}
\bar{D}_{\Delta_{1}\Delta_{2}\Delta_{3}\Delta_{4}}(u,v)=
\bar{D}_{\Delta_{1}\Delta_{2}\Delta_{3}\Delta_{4}}(u,v)_{\mathrm{reg}}+
\bar{D}_{\Delta_{1}\Delta_{2}\Delta_{3}\Delta_{4}}(u,v)_{\mathrm{sing}}
\end{equation}
with the regular part is of order $O(u^m,u^m\log{u})$ for $m\geq 0$ and the singular part is given by
\begin{eqnarray}
\bar{D}_{\Delta_{1}\Delta_{2}\Delta_{3}\Delta_{4}}(u,v)_{\mathrm{sing}}&=&u^{-s}
\frac{\Gamma(\Delta_{1}-s)\Gamma(\Delta_{2}-s)\Gamma(\Delta_{3})
\Gamma(\Delta_{4})}{\Gamma(\Delta_{3}+\Delta_{4})} \nonumber \\
&\times& \sum_{m=0}^{s-1}(-1)^{m}(s-m-1)!\frac{(\Delta_{1}-s)_{m}(\Delta_{2}-s)_{m}
(\Delta_{3})_{m}(\Delta_{4})_{m}}{m!(\Delta_{3}+\Delta_{4})_{2m}} \nonumber \\
&\times& u^{m}F(\Delta_{2}-s+m,\Delta_{3}+m, \Delta_{3}+\Delta_{4}+2m;1-v)
\label{Dsingular}
\end{eqnarray}
where we have introduced
\begin{equation}
s=\frac{1}{2}\left(\Delta_{1}+\Delta_{2}-\Delta_{3}-\Delta_{4}\right)
\end{equation}

The regular part contains terms of the form $\ln u$ that lead to contributions to anomalous dimensions of order $1/N^2$ coming from long operators.  When discussing the case of $[0,p,0]$ chiral primaries, long operators belong to $SU(4)$ representations $[n-m,2m,n-m]$ satisfying $m \leq n \leq p-2$ \cite{Eden:2001ec, Heslop:2001gp}. Previous computations have indicated that at strong coupling, the only operators that develop anomalous dimensions have $\Delta-l \geq 2p$, which implies that long multiplets with twist $\Delta-l < 2p$ are absent in the OPE of two CPOs of weight $p$ in the large $N$ limit. 

In \cite{Arutyunov:2000ku} it was shown that for the $p=2$ case, the only twist 2 operator appearing in the OPE at large $N$ is the energy momentum tensor. It was confirmed in \cite{Dolan:2001tt} that all other twist two operators are absent at large $N$ for any $l$, which differs to what happens at weak coupling, where twist 2 operators appear for any $l$ (with $l=0$ corresponding to the Konishi scalar). The absence of twist 2 operators when the theory becomes strongly coupled requires a non-trivial cancellation of between free field theory contributions to the OPE and dynamical contributions arising from the non-logarithmic regular part of  the $\bar{D}$-functions.  This is, the sub-leading terms from the $\bar{D}$-function which correspond to operators with $\Delta-l < p$ cancel with the corresponding free field contributions.

This cancellation was exploited by Dolan, Nirschl and Osborn in \cite{Dolan:2006ec} to propose the form of the strongly coupled amplitude. In this paper we have shown from supergravity that next-next-to-extremal correlators can be reduced to a single $\bar{D}$-function of the conjectured from\footnote{The cases $p=3,4$ have also shown to agree in \cite{Dolan:2004iy}.}. The presence of mixing between contributions coming from the dynamical and free parts of the amplitude suggests that there is no reason to expect that $\mathcal{G}_{0}(u,v;\sigma,\tau)$ will precisely match the ``free'' part of $\mathcal{G}^{\mathrm{sugra}}(u,v;\sigma,\tau)$ and as it happens, our result (\ref{ressc}) differs from the free theory result (\ref{freeresultamp}). The same phenomena was observed in \cite{Uruchurtu:2008kp}.

Given these arguments, it would be interesting to analyse the contributions coming from different multiplets to the partial wave expansion and to verify explicitly the cancelations between the free field values and the results from supergravity. Of interest as well would be to try to generalise the conjecture in  \cite{Dolan:2006ec} to all four point functions of $1/2$-BPS chiral primary operators, namely, correlators of the form $\langle \mathcal{O}_{p_1}\mathcal{O}_{p_2}\mathcal{O}_{p_3}\mathcal{O}_{p_4}\rangle$. From our results it is straightforward to conjecture the strongly coupled limit of $\mathcal{G}(u,v;\sigma,\tau)$ for the most generic next-next-to-extremal case\footnote{Large $N$ free field computations such as the ones in section \ref{sec:FreeFieldTh} were carried out to work out the coefficients.}. This is
\begin{align}
\mathcal{G}(u,v;\sigma,\tau)&=\frac{\sqrt{p_1p_2p_3p_4}}{N^2}\Big\{  (p_1-1)\sigma u +(p_2-1)\tau\frac{u}{v}+(p_3-1)\sigma\tau\frac{u^2}{v}
\nonumber \\
&-\left.\frac{1}{\Gamma(p_1-1)\Gamma(p_2-1)\Gamma(p_3-1)}s(u,v;\sigma, \tau) u^{p_3} v^{p_{1}-2} \bar{D}_{p_3 \ p_4+2 \ p_1\ p_2}(u,v)\right\}
\end{align}

In a recent paper \cite{Buchbinder:2010ek}, Buchbinder and Tseytlin discussed semi-classical methods for evaluating four point amplitudes of classical strings. In particular, they focused on two ``heavy'' vertex operators with large quantum numbers and two ``light'' operators and argued the result can be written as a product of two three point functions. For the case in which the operators were taken to be chiral primary scalars of charges $\pm J$ and $\pm j$ with $J \gg j$, their result reads
\begin{equation}
\langle V_{-J}(\vec{x}_{1}) V_{J}(\vec{x}_{2}) V_{-j}(\vec{x}_{3}) V_{j}(\vec{x}_{4})   \rangle = \frac{\mathcal{F}(u,v)}{\vert \vec{x}_{12} \vert^{2J}\vert \vec{x}_{34} \vert^{2j}}, \qquad
\qquad \mathcal{F}(u,v)=\frac{jJ^{2}}{N^{2}}\frac{u^{j}}{v^{j}}
\label{ressc}
\end{equation}
This agrees with the free field SYM theory result in the limit in which $J \geq 1$ \cite{Rayson:2007th}. An attempt to compare (\ref{ressc}) with the supergravity result in \cite{Uruchurtu:2008kp} uncovered an apparent disagreement, as the supergravity result had no term of order $J^{2}$. There it was argued that the dynamical function $\mathcal{H}_{I}(u,v)$\footnote{In \cite{Buchbinder:2010ek}, $\mathcal{H}_{I}(u,v)$ was denoted $\hat{D}(u,v,J)$.} was of order $J^{1/2}$, hence sub-leading at large $J$, and it was suggested that there could be some missing terms.

A careful analysis using the short-distance expansion in (\ref{Dsingular}) reveals that as it happens, $\mathcal{H}_{I}(u,v)$  is $O(J)$ as $u \rightarrow 0$, comparable to the contributions coming from the ``free'' part. This behaviour is consistent with the arguments presented above, since contributions coming from the dynamical part should mix with those from the free part. These same arguments makes us confident in our result (\ref{ampfilnal}). A more detailed analysis taking into account the short distance expansion expressions above might reveal terms of order $J^{2}$. Also it would be necessary to question the order of limits taken in  \cite{Buchbinder:2010ek}, specifically the short-distance expansions in $u$,$v$ and the limit in which $J$ becomes large. We hope to address these questions in the future.

\appendix
\section{Harmonic Variables and Propagator Basis}
\label{sec:harmpoly}
The quartic couplings are described in terms of products of two Clebsch-Gordan coefficients
\begin{equation}
\inp{C_{k_1}^1 C_{k_2}^2 C_{[a_1, a_2, a_3]}^I}\inp{C_{k_3}^3 C_{k_4}^4 C_{[a_1, a_2, a_3]}^I},
\label{eq:clebschsum}
\end{equation}
that arise from overlapping integrals of spherical harmonics when compactifying the supergravity action on $S^5$.

In order to compare the supergravity induced four point function with the one computed in field theory, we need  to expand this rank 4 tensors in terms of the so-called propagator basis. The basis in the case $k_1=k_2=m$ and $k_3=k_4=n$ is given by 6 possible $SO(6)$ tensors which have the form
\begin{align}
I^{1234}&=C^{1}_{ij l_{1}\cdots l_{k}}C^{2}_{i j m_{1}\cdots m_{k}}C^{3}_{p_{1}\cdots p_{n-k}}C^{4}_{p_{1}\cdots p_{n-k}l_{1}\cdots l_{k}m_{1}\cdots m_{k}} \nonumber \\
C^{1234}&=C^{1}_{ijl_{1}\cdots l_{k}}C^{2}_{jpm_{1}\cdots m_{k}}C^{3}_{pq_{1}\cdots q_{n-k-1}}C^{4}_{i q_{1}\cdots q_{n-k-1}l_{1}\cdots l_{k}m_{1}\cdots m_{k}} \nonumber \\
C^{2143}&=C^{1}_{ijl_{1}\cdots l_{k}}C^{2}_{pi m_{1}\cdots m_{k}}C^{3}_{j q_{1}\cdots q_{n-k-1}}C^{4}_{p q_{1}\cdots q_{n-k-1}l_{1}\cdots l_{k}m_{1}\cdots m_{k}} \nonumber \\
\Upsilon^{1234}&=C^{1}_{ijl_{1}\cdots l_{k}}C^{2}_{m_{1}\cdots m_{k+2}}C^{3}_{ij q_{1}\cdots q_{n-k-2}}C^{4}_{q_{1}\cdots q_{n-k-2}l_{1}\cdots l_{k}m_{1}\cdots m_{k+2}} \nonumber \\
\Upsilon^{2143}&=C^{1}_{l_{1}\cdots l_{k+2}}C^{2}_{pr m_{1}\cdots m_{k}}C^{3}_{pr q_{1}\cdots q_{n-k-2}}C^{4}_{q_{1}\cdots q_{n-k-2}l_{1}\cdots l_{k+2}m_{1}\cdots m_{k}} \nonumber \\
S^{1234}&=C^{1}_{ip l_{1}\cdots l_{k}}C^{2}_{j r m_{1}\cdots m_{k}}C^{3}_{ r pq_{1}\cdots q_{n-k-2}}C^{4}_{ijq_{1}\cdots q_{n-k-2}l_{1}\cdots l_{k}m_{1}\cdots m_{k}} 
\end{align}
Each of these correspond to one of the diagrams in figure \ref{colourbasis}. The definitions above imply the following symmetry properties:
\begin{equation}
C^{1234}=C^{2143} \qquad \qquad \Upsilon^{1234}=\Upsilon^{2143}
\end{equation}
As it was shown in section (\ref{sec:structure}), one can also express four point amplitudes in terms of a basis for $SO(6)$ tensor fields by introducing six-dimensional complex null vectors $t_i$, so the correlator becomes an invariant function of the $SU(4)$ variables $\sigma$ and $\tau$ introduced in (\ref{sigmatau}).
We can then re-express the rank 4 tensors in terms of harmonic variables. In the $s$-channel (1234) we have:
\begin{align}
 \Upsilon^{1234}&\rightarrow  \sigma^{2} \qquad \Upsilon^{2134} \rightarrow \tau^{2} \nonumber \\
C^{1234} &\rightarrow   \sigma \qquad C^{2134} \rightarrow \tau \nonumber \\
S^{1234}  &\rightarrow \sigma\tau \qquad  I^{1234} \rightarrow 1 \nonumber \\
\end{align}
For the remaining channels, one has the transformations
\begin{equation}
\tilde \sigma \rightarrow \frac{1}{\sigma} \qquad \qquad \tilde \tau \rightarrow \frac{\tau}{\sigma}
\end{equation}
for the $t$-channel (1324) and for the $u$-channel (1423)
\begin{equation}
\hat \sigma \rightarrow \frac{\sigma}{\tau} \qquad \qquad \hat \tau \rightarrow \frac{1}{\tau}
\end{equation}
\section{Integration over $S^5$}
\label{sec:integrationsphere}
Define spherical harmonics transforming in $[0,k,0]$ irrep. of $SU(4)$ as 
\begin{equation}
Y^I_k = z(k) C^I_{i_1...i_k} \xi^{i_1}... \xi^{i_k}
\end{equation}
where $\xi \in S^5$, i.e. $\xi^2=1$. The basis of totally symmetric traceless tensors $C^I_{i_1...i_k}$ is orthonormal
\begin{equation} 
C^{I_1}_{i_1...i_k}C^{I_2}_{i_1...i_k}=\delta^{I_1I_2}.
\end{equation}
and
\begin{equation}
z(k)= \sqrt{2^{k-1}(k+1)(k+2)}.
\end{equation}
It follows that
\begin{equation}
\int_{S^5}Y^{I_1}_k Y^{I_2}_k = \omega_5 \delta^{I_1I_2}.
\end{equation}
From \cite{Osborn:2010sp} we have that
\begin{equation}
\int_{S^5}(t_1 \cdot \xi)^k (t_2 \cdot \xi)^k = \frac{\omega_5}{2^{k-1}(k+1)(k+2)} (t_1 \cdot t_2)^k=\frac{\omega_5}{(z(k))^2}(t_1 \cdot t_2)^k
\end{equation}
Hence 
\begin{equation}
Y^{(k)}_k \rightarrow z(k) (t \cdot \xi)^k.
\end{equation}
We are interested in evaluating the integral
\begin{align}
& \int_{S^5} d \Omega_1\int_{S^5}d \Omega_2 (t_1 \cdot \xi_1)^{k_1} (t_2 \cdot \xi_1)^{k_2} \sum_{I_5} Y^{I_5}_{k_5}(\xi_1) Y^{I_5}_{k_5}(\xi_2) (t_3 \cdot \xi_2)^{k_3} (t_4 \cdot \xi_2)^{k_4}
\end{align}
assuming $k_4 \geq k_3$ and $a,b \geq 0$, the result has been obtained in \cite{Osborn:2010sp} and we reproduce it below
\begin{eqnarray*}
& \int_{S^5} d \Omega_1\int_{S^5}d \Omega_2 (t_1 \cdot \xi_1)^{k_1} (t_2 \cdot \xi_1)^{k_2} \sum_{I_5} Y^{I_5}_{k_5}(\xi_1) Y^{I_5}_{k_5}(\xi_2) (t_3 \cdot \xi_2)^{k_3} (t_4 \cdot \xi_2)^{k_4}\\
& = \frac{\omega_5^2}{2^{\Sigma-1}} \frac{k_1! k_2! k_3! k_4! (k_5+2)}{(\sigma_{125}+2)! (\sigma_{345}+2)! \alpha_{125}! \alpha_{345}!} F^{k_5}_{k_2-k_1,k_4-k_3}(\sigma,\tau) (t_1 \cdot t_2)^{\Sigma-k_4} (t_1 \cdot t_2)^{k_3} (t_1 \cdot t_2)^a (t_1 \cdot t_2)^b
\end{eqnarray*}
where
\begin{align*}
\Sigma &= \frac{1}{2}(k_1+k_2+k_3+k_4),\\
\; a &= \frac{1}{2}(k_1+k_4-k_2-k_3), \; \qquad \qquad b=\frac{1}{2}(k_2+k_4-k_1-k_3),\\
\; \sigma_{ijl} &= \frac{1}{2}(k_i+k_j+k_l), \; \qquad \qquad \qquad \alpha_{ijl}=\frac{1}{2}(k_i+k_j-k_l),
\end{align*}
and 
\begin{equation}
F^{(a+b+2n)}_{b-a,a+b}(\sigma,\tau)=\frac{(a+b+2n+1)!}{a!b!}Y^{(a,b)}_{nn}(\sigma,\tau)
\end{equation}
with
\begin{equation}
a+b+2n=k_5 \quad \Rightarrow \quad n=\alpha_{534}.
\end{equation}
From \cite{Arutyunov:1999en} we have\footnote{Note the inclusion of a factor of $\omega_5^{3/2}$ w.r.t. that paper, since the normalisation of the spherical harmonics that they use includes an additional $\omega_5^{1/2}$.}
\begin{equation}
a_{I_1I_2I_3} \equiv \omega_5^{3/2}\int_{S^5} Y^{I_1}_{k_1}Y^{I_2}_{k_2}Y^{I_3}_{k_3} \quad = A_{123}(k_1,k_2,k_3) \langle C^{I_1}_{[0,k_1,0]}C^{I_2}_{[0,k_2,0]}C^{I_3}_{[0,k_3,0]}\rangle\ 
\end{equation}
where 
\begin{equation}
A_{123}(k_1,k_2,k_3) = \frac{\omega_5}{2^{\sigma_{123}-1}}\frac{k_1! k_2! k_3! z(k_1) z(k_2) z(k_3)}{(\sigma_{123}+2)!\alpha_{231}! \alpha_{132}! \alpha_{123}!}
\end{equation}
and  
\begin{equation}
\langle C_{[0,k_{1},0]}^{I_{1}}C_{[0,k_{2},0]}^{I_{2}}C_{[0,k_{3},0]}^{I}\rangle=C^{I_{1}}_{i_{1}...i_{\alpha_{2}}j_{1}...j_{\alpha_{3}}}
C^{I_{2}}_{j_{1}...j_{\alpha_{3}}l_{1}...l_{\alpha_{1}}}C^{I}_{l_{1}...l_{\alpha_{1}}i_{1}...i_{\alpha_{2}}}
\end{equation}
and $\alpha_k=\alpha_{ijk}$ for $j \neq k \neq i$. Using the identifications above, it follows that, for fixed $k_5$ and for the highest weight components of the representations
\begin{align*}
a_{I_1 I_2I }a_{I_3I_4I} &\equiv \omega_5^3 \int_{S^5} \int_{S^5} Y^{I_1}_{k_1}Y^{I_2}_{k_2}\sum_{I}Y^{I}_{k_5} Y^{I_5}_{k_5} Y^{I_3}_{k_3}Y^{I_4}_{k_4}  \\
& \simeq z(k_1) z(k_2) z(k_3) z(k_4)  \int_{S^5} d \Omega_1\int_{S^5}d \Omega_2 (t_1 \cdot \xi_1)^{k_1} (t_2 \cdot \xi_1)^{k_2} \sum_{I} Y^{I}_{k_5}(\xi_1) Y^{I}_{k_5}(\xi_2) (t_3 \cdot \xi_2)^{k_3} (t_4 \cdot \xi_2)^{k_4} 
\end{align*}
Substituting the expressions for the integrals we get
\begin{align*}
& \langle C^{I_1}_{[0,k_1,0]}C^{I_2}_{[0,k_2,0]}C^{I}_{[0,k_5,0]}\rangle \langle C^{I_3}_{[0,k_3,0]}C^{I_4}_{[0,k_4,0]}C^{I}_{[0,k_5,0]}\rangle  \\
& \rightarrow \frac{ (t_1 \cdot t_2)^{\Sigma-k_4} (t_1 \cdot t_2)^{k_3} (t_1 \cdot t_2)^a (t_1 \cdot t_2)^b}{A_{125}(k_1,k_2,k_5)A_{345}(k_3,k_4,k_5)} \frac{z(k_1) z(k_2) z(k_3) z(k_4)}{2^{\Sigma-1}} \frac{k_1! k_2! k_3! k_4!  (k_5+2)}{(\sigma_{125}+2)! (\sigma_{345}+2)! \alpha_{125}! \alpha_{345}!} \frac{(k_5+1)!}{a!b!}Y^{(a,b)}_{nn} \times \\
& = \frac{2^{\sigma_{125}-1} 2^{\sigma_{345}-1} \alpha_{251}! \alpha_{512}! \alpha_{453}! \alpha_{534}!}{(k_5!)^2 z(k_5)^2} \frac{1}{2^{\Sigma-1}}\frac{(k_5+2)!}{a!b!}Y^{(a,b)}_{nn}  (t_1 \cdot t_2)^{\Sigma-k_4} (t_1 \cdot t_2)^{k_3} (t_1 \cdot t_2)^a (t_1 \cdot t_2)^b \\
& = \frac{ \alpha_{251}! \alpha_{512}! \alpha_{453}! \alpha_{534}!}{k_5!} \frac{1}{a!b!}Y^{(a,b)}_{nn}  (t_1 \cdot t_2)^{\Sigma-k_4} (t_1 \cdot t_2)^{k_3} (t_1 \cdot t_2)^a (t_1 \cdot t_2)^b
\end{align*}
where in the last equality $\sigma_{125}+\sigma_{345}-\Sigma=k_5$ was used. This normalisation agrees with that in \cite{Berdichevsky:2007xd} and \cite{Uruchurtu:2008kp}. The same procedure can be followed for integrals involving vector spherical harmonics and tensor harmonics
\begin{align}
t_{123}&=\omega_{5}^{3/2}\int \nabla^{\alpha}Y^{I_{1}}_{k_{1}}Y^{I_{2}}_{k_{2}}Y^{I_3}_{ \alpha k_{3}}=T_{123}(k_{1},k_{2},k_{3})
\langle C_{[0,k_{1},0]}^{I_1}C_{[0,k_{2},0]}^{I_2}C_{[1,k_{3}-1,1]}^{I_3}\rangle \nonumber \\
p_{123}&=\omega_{5}^{3/2}\int \nabla^{\alpha}Y^{I_{1}}_{k_{1}}\nabla^{\beta}Y^{I_{2}}_{k_{2}}Y^{I_3}_{( \alpha \beta)k_{3}}=P_{123}(k_{1},k_{2},k_{3})
\langle C_{[0,k_{1},0]}^{I_1}C_{[0,k_{2},0]}^{I_2}C_{[2,k_{3}-2,2]}^{I_3}\rangle
\end{align}
where
\begin{align}
\langle C_{[0,k_{1},0]}^{I_1}C_{[0,k_{2},0]}^{I_2}C_{[1,k_{3}-1,1]}^{I}\rangle&=C^{I_{1}}_{mi_{1}...i_{p_{2}}j_{1}...j_{p_{3}}}
C^{I_{2}}_{j_{1}...j_{p_{3}}l_{1}...l_{p_{1}}}C^{I}_{m;l_{1}...l_{p_{1}}i_{1}...i_{p_{2}}}
\nonumber \\
&-C^{I_{1}}_{i_{1}...i_{p_{2}+1}j_{1}...j_{p_{3}}}C^{I_{2}}_{j_{1}...j_{p_{3}}l_{1}...l_{p_{1}-1}}
C^{I}_{m;l_{1}...l_{p_{1}-1}i_{1}...i_{p_{2}+1}}
\end{align}
where $p_{1}=\alpha_{1}+\frac{1}{2}$, $p_{2}=\alpha_{2}-\frac{1}{2}$ and $p_{3}=\alpha_{3}-\frac{1}{2}$, and
\begin{equation}
\langle C_{[0,k_{1},0]}^{I_1}C_{[0,k_{2},0]}^{I_2}C_{[2,k_{3}-2,2]}^{I}\rangle=C^{I_{1}}_{mi_{1}...i_{p_{2}}j_{1}...j_{p_{3}}}
C^{I_{2}}_{nj_{1}...j_{p_{3}}l_{1}...l_{p_{1}}}C^{I}_{mn;l_{1}...l_{p_{1}}i_{1}...i_{p_{2}}}
\end{equation}
We list the results below. The normalisation constant relating $Y_{nn}^{(a,b)}$ to $\langle C^{I_1}_{k_1}C^{I_2}_{k_2}C^I_{[0,n-m+2k,0]}\rangle \langle C^{I_3}_{k_3}C^{I_4}_{k_4}C^I_{[0,n-m+2k,0]}\rangle$ is given by
\begin{equation}
\frac{\Gamma \left(\frac{1}{2} ({k_1}-{k_2}+{k_5}+2)\right) \Gamma \left(\frac{1}{2} (-{k_1}+{k_2}+{k_5}+2)\right) \Gamma
   \left(\frac{1}{2} ({k_3}-{k_4}+{k_5}+2)\right) \Gamma \left(\frac{1}{2} (-{k_3}+{k_4}+{k_5}+2)\right)}{\Gamma ({k_5}+1) \Gamma
   \left(\frac{1}{2} ({k_1}-{k_2}-{k_3}+{k_4}+2)\right) \Gamma \left(\frac{1}{2} (-{k_1}+{k_2}-{k_3}+{k_4}+2)\right)}
\end{equation}
The normalisation constant relating $Y_{nn-1}^{(a,b)}$ to  $\langle C^{I_1}_{k_1}C^{I_2}_{k_2}C^I_{[1,n-m+2k-1,0]}\rangle \langle C^{I_3}_{k_3}C^{I_4}_{k_4}C^I_{[1,n-m+2k-1,1]}\rangle$ reads
\begin{equation}
\frac{\Gamma ({k_5}+2) \Gamma \left(\frac{1}{2} ({k_1}-{k_2}+{k_5}+1)\right) \Gamma \left(\frac{1}{2} (-{k_1}+{k_2}+{k_5}+1)\right)
   \Gamma \left(\frac{1}{2} ({k_3}-{k_4}+{k_5}+1)\right) \Gamma \left(\frac{1}{2} (-{k_3}+{k_4}+{k_5}+1)\right)}{({k_5}!)^2
   \left(\frac{1}{2} ({k_1}-{k_2}-{k_3}+{k_4})\right)! \left(\frac{1}{2} (-{k_1}+{k_2}-{k_3}+{k_4})\right)!}
\end{equation}
Finally, the normalisation constant relating $Y_{nn-2}^{(a,b)}$ to $\langle C^{I_1}_{k_1}C^{I_2}_{k_2}C^I_{[2,n-m+2k-2,2]}\rangle \langle C^{I_3}_{k_3}C^{I_4}_{k_4}C^I_{[2,n-m+2k-2,2]}\rangle$
\begin{equation}
\frac{16 \Gamma \left(\frac{1}{2} ({k_1}-{k_2}+{k_5}+2)\right) \Gamma \left(\frac{1}{2} (-{k_1}+{k_2}+{k_5}+2)\right) \Gamma
   \left(\frac{1}{2} ({k_3}-{k_4}+{k_5}+2)\right) \Gamma \left(\frac{1}{2} (-{k_3}+{k_4}+{k_5}+2)\right)}{{k_5} \Gamma ({k_5}+2)
   \Gamma \left(\frac{1}{2} ({k_1}-{k_2}-{k_3}+{k_4}+2)\right) \Gamma \left(\frac{1}{2} (-{k_1}+{k_2}-{k_3}+{k_4}+2)\right)}
\end{equation}

It is also convenient to introduce the relation between the harmonic polynomials  $Y_{nm}^{(a,b)}(\sigma,\tau)$ and combinations of Jacobi polynomials \cite{Rayson:2007th}. Let us start by introducing
\begin{equation}
P_{nm}^{(a,b)}(y,\bar y)=\frac{P_{n+1}^{(a,b)}(y)P_{m}^{(a,b)}(\bar y)-P_{m}^{(a,b)}(y)P_{n+1}^{(a,b)}(\bar y)}{y-\bar y}
\end{equation}
where $y$ and $\bar{y}$ are related to $\sigma$ and $\tau$ by
\begin{equation}
\sigma=\frac{1}{4}(y+1)(\bar{y}+1) \qquad \qquad \tau=\frac{1}{4}(1-y)(1-\bar{y})
\end{equation}
and $P_{n}^{(a,b)}(y)$ are the standard Jacobi polynomials. For $m=n$ one has that
\begin{equation}
Y_{nn}^{(a,b)}=\frac{2 (n+1)! (a+b+n+1)!}{(a+1)_n (b+1)_n (a+b+2 n+2)!}P_{nn}^{(a,b)}
\end{equation}
For $m=n-1$
\begin{equation}
Y_{nn-1}^{(a,b)}=\frac{2 (n+1)! (a+b+n+1)!}{(a+1)_{n-1} (b+1)_{n-1} (a+b+2 n+2)!}P_{nn-1}^{(a,b)}
\end{equation}
and finally, for $m=n-2$
\begin{equation}
Y_{nn-2}^{(a,b)}= \frac{2 (n+1)! (a+b+n+1)!}{(a+1)_{n-2} (b+1)_{n-2} (a+b+2 n+2)!} P_{nn-2}^{(a,b)}
\end{equation}
Using these relations we can obtain the explicit form of the effective couplings arising from the integration on $S^{5}$ for our particular process.  Consider first the products of scalar harmonics $a_{125}a_{345}$ for fixed $k_{5}$.  In the $s$-channel the relevant polynomials are
\begin{align}
Y_{00}^{(k,k)}&=1 \nonumber \\
Y_{11}^{(k,k)}&=\frac{\sigma}{k+1}+\frac{\tau}{k+1}-\frac{1}{2k+3} \nonumber \\
Y_{22}^{(k,k)}&=\frac{\sigma^{2}+\tau^{2}}{2(k+1)(k+2)}+\frac{\sigma\tau}{(k+1)^{2}}-\frac{\sigma+\tau}{(k+1)(2k+5)}+\frac{1}{2(2k+4)(2k+5)}
\label{124eq}
\end{align}
which correspond to $k_{5}=2k,2k+2,2k+4$. In the $t$-channel, the polynomials that we will use are
\begin{align}
Y_{00}^{k,n-k-2}&=1 \nonumber \\
Y_{11}^{(k,n-k-2)}&=\frac{\sigma}{n-k-1}+\frac{\tau}{k+1}-\frac{1}{n+1} \nonumber \\
Y_{22}^{(k,n-k-2)}&=\frac{\sigma^{2}}{2(n-k-1)(n-k)}+\frac{\sigma\tau}{(k+1)(n-k-1)}-\frac{\sigma}{(n-k-1)(n+3)}
\nonumber \\
&-\frac{\tau}{(k+1)(n+3)}+\frac{\tau^{2}}{2(k+1)(k+2)}+\frac{1}{2(n+2)(n+3)}
\label{125eq}
\end{align}
which equates to the cases $k_{5}=n-2,n,n+2$. For products of vector harmonics $t_{125}t_{345}$, in the $s$-channel, we will make use of 
\begin{align}
Y_{10}^{(k,k)}&=\frac{2 (k+1) }{2k+1}(\sigma -\tau ) \nonumber \\
Y_{21}^{(k,k)}&=\frac{(k+2)}{2(2k+3)}(2\sigma^2-2\tau^2-\sigma+\tau) 
\end{align}
corresponding to $k_5=2k+1, 2k+3$ and of
\begin{align}
Y_{10}^{(k,n-k-2)}&=\frac{n (-(n-2k-2) +(n+2) )}{(n+2)(n-1)}(\sigma -\tau )
\nonumber \\
Y_{21}^{(k,n-k-2)}&=-\frac{(k+1) (n+2) (n-k-1)}{n (n+1)}\Big[ \frac{-n+2k+2}{(n+4)(n+2)}+\frac{n-3k-5}{(k+1) (n+4)}\sigma+\frac{2n-3k-1}{(n-k-1) (n+4)}\tau
\nonumber \\
&+\frac{\sigma^2}{k+1}-\frac{\tau^2}{n-k-1}-\frac{n-2k-2}{(k+1)(n-k-1)}\sigma\tau \Big]
\end{align}
for the $t$-channel, for which $k_5=n-1,n+1$.

The only relevant polynomial for the product of tensor harmonics $p_{125}p_{345}$ happens to be $Y_{20}$. In the $s$-channel, $Y_{20}^{(k,k)}$ and $k_5=2k+2$ whereas in the $t$-channel, $Y_{20}^{(k,n-k-2)}$ for which $k_5=n$. Explicitly
\begin{align}
Y_{20}^{(k,k)}&= \frac{2(k+1)}{(2k+3)(2k+5)}+\frac{2}{(2k+3)}\left(2 \sigma ^2+2\tau^{2}-\sigma  -\tau -4 \sigma\tau\right)
\nonumber \\
Y_{20}^{(k,n-k-2)}&=\frac{16 (k+1) (n-k-1) }{n^2 (n+1) (n+3) (n+4)}\Big[ \left(3 (k+1)^2-(3 k+2) n+n^2\right) +(n+3)(n-3k-5)\sigma +(n+3)(n+4) \sigma^{2} 
\nonumber \\
&-(2n-3k-1)\tau -2(n+4) \sigma\tau +(n+4) \tau ^2\Big] 
\end{align}
\section{Properties of $D$-Functions}
\label{sec:Dfunc}
We collect here some general properties and identities of the $D$-functions.  These are defined as integrals over $AdS_{5}$, by the formula
\begin{equation}
D_{\Delta_{1}\Delta_{2}\Delta_{3}\Delta_{4}}(\vec{x}_{1},\vec{x}_{2},\vec{x}_{3},\vec{x}_{4})=\int
\frac{d^{5}z}{z_{0}^{5}}\tilde{K}_{\Delta_{1}}(z,\vec{x}_{1})\tilde{K}_{\Delta_{2}}(z,\vec{x}_{2})
\tilde{K}_{\Delta_{3}}(z,\vec{x}_{3})\tilde{K}_{\Delta_{4}}(z,\vec{x}_{4})
\end{equation}
with
\begin{equation}
\tilde{K}_{\Delta}(z,\vec{x})=\left(\frac{z_{0}}{z_{0}^2+(\vec{z}-\vec{x})^{2}}\right)^{\Delta}
\label{Kpropapp}
\end{equation}
$\bar{D}$-functions, which are functions of conformal invariant ratios $u$ and $v$, can be defined from $D$-functions by
\begin{equation}
\bar{D}_{\Delta_{1}\Delta_{2}\Delta_{3}\Delta_{4}}(u,v)=\kappa
\frac{|\vec{x}_{31}|^{2\Sigma-2\Delta_{4}}|\vec{x}_{24}|^{2\Delta_{2}}}
{|\vec{x}_{41}|^{2\Sigma-2\Delta_{1}-2\Delta_{4}}|\vec{x}_{34}|^{2\Sigma-2\Delta_{3}-2\Delta_{4}}}
D_{\Delta_{1}\Delta_{2}\Delta_{3}\Delta_{4}}(\vec{x}_1,\vec{x}_2,\vec{x}_3,\vec{x}_4)
\label{Dbardef}
\end{equation}
where $2\Sigma=\sum_{i}\Delta_{i}$ and
\begin{equation}
\kappa=\frac{2}{\pi^2}\frac{\Gamma(\Delta_{1})\Gamma(\Delta_{2})\Gamma(\Delta_{3})\Gamma(\Delta_{4})}{\Gamma(\Sigma-2)}
\end{equation}
Some identities relating $\bar{D}$-functions with different values of $\Sigma$ are listed as follows:
\begin{align}
(\Delta_{2}+\Delta_{4}-\Sigma)\bar{D}_{\Delta_{1}\Delta_{2}\Delta_{3}\Delta_{4}}
&=\bar{D}_{\Delta_{1}\Delta_{2}+1\Delta_{3}\Delta_{4}+1}
-\bar{D}_{\Delta_{1}+1\Delta_{2}\Delta_{3}+1\Delta_{4}} \nonumber \\
(\Delta_{1}+\Delta_{4}-\Sigma)\bar{D}_{\Delta_{1}\Delta_{2}\Delta_{3}\Delta_{4}}
&=\bar{D}_{\Delta_{1}+1\Delta_{2}\Delta_{3}\Delta_{4}+1}
-v\bar{D}_{\Delta_{1}\Delta_{2}+1\Delta_{3}+1\Delta_{4}} \nonumber \\
(\Delta_{3}+\Delta_{4}-\Sigma)\bar{D}_{\Delta_{1}\Delta_{2}\Delta_{3}\Delta_{4}}
&=\bar{D}_{\Delta_{1}\Delta_{2}\Delta_{3}+1\Delta_{4}+1}
-u\bar{D}_{\Delta_{1}+1\Delta_{2}+1\Delta_{3}\Delta_{4}}
\label{usefulDid1}
\end{align}
Furthermore, there are identities relating $\bar{D}$-functions with the same $\Sigma$. The most frequently used in manipulations throughout this paper is
\begin{align}
\Delta_{4}\bar{D}_{\Delta_{1}\Delta_{2}\Delta_{3}\Delta_{4}}=\bar{D}_{\Delta_{1}\Delta_{2}\Delta_{3}+1\Delta_{4}+1}
+\bar{D}_{\Delta_{1}\Delta_{2}+1\Delta_{3}\Delta_{4}+1}
+\bar{D}_{\Delta_{1}+1\Delta_{2}\Delta_{3}\Delta_{4}+1}
\label{usefulDid2}
\end{align}
A useful expression arises when we take the limit  of (\ref{usefulDid2})  in which one of the $\Delta_{i}\rightarrow 0$ to obtain
\begin{align}
\bar{D}_{\Delta_{1}\Delta_{2}\Delta_{3}+1\Delta_{4}+1}
+u\bar{D}_{\Delta_{1}+1\Delta_{2}+1\Delta_{3}\Delta_{4}}
+v\bar{D}_{\Delta_{1}\Delta_{2}+1 \Delta_{3}+1 \Delta_{4}}\vert_{\Delta_{1}+\Delta_{2}+\Delta_{3}=\Delta_{4}}
\nonumber \\
=\Gamma(\Delta_{1})\Gamma(\Delta_{2})\Gamma(\Delta_{3})
\label{combidelta}
\end{align}
Further properties of these functions can be found in Appendix C of  \cite{Uruchurtu:2008kp}. 
\section{Various Manipulations involving $\bar{D}$-functions}
\label{sec:Dfuncsimp}
Here we show how to simplify the final supergravity result (\ref{resultdiffweightsugra}) to show that x is equivalent to y. All $\bar{D}$-functions entering (\ref{along})-(\ref{dlong}), have $\Sigma=n+k+3, n+k+2$ and $n+k+1$. We start by showing that all $\bar{D}$-functions with $\Sigma=n+k+3$ can be expressed in terms of $\bar{D}_{k+2 \ k+2 \ n-k \ n+k+2}$. We then show that $\bar{D}$-functions of lower $\Sigma$ cancel each other out.  

Consider  $\tilde{a}(u,v)$. We start by using the identities (\ref{usefulDid1}) on the terms with $\Sigma=n+k+3$. This procedure generates new $\bar{D}$-terms with equal $\Sigma$ and terms with $\Sigma=n+k+2$
\begin{align}
v\bar{D}_{k+1 \ k+3 \ n-k+1 \ n+k+1}&=\bar{D}_{k+2 \ k+2 \ n-k \ n+k+2}-k\bar{D}_{k+1 \ k+2 \ n-k \ n+k+1} \nonumber \\
\bar{D}_{k+3 \ k+1 \ n-k+1 \ n+k+1}&=\bar{D}_{k+2 \ k+2 \ n-k \ n+k+2}-k\bar{D}_{k+2 \ k+1 \ n-k \ n+k+1} \nonumber \\
\bar{D}_{k+2 \ k+2 \ n-k+1 \ n+k+1}&=\bar{D}_{k+1 \ k+3 \ n-k \ n+k+2}-(k+1)\bar{D}_{k+1 \ k+2 \ n-k \ n+k+1} \nonumber \\
v\bar{D}_{k+2 \ k+2 \ n-k+1 \ n+k+1}&=\bar{D}_{k+3 \ k+1 \ n-k \ n+k+2}-(k+1)\bar{D}_{k+2 \ k+1 \ n-k \ n+k+1} \nonumber \\
u\bar{D}_{k+2 \ k+2 \ n-k+1 \ n+k+1}&=\bar{D}_{k+1 \ k+1 \ n-k-2 \ n+k+2}-(n-k)\bar{D}_{k+1 \ k+1 \ n-k+1 \ n+k+1} \nonumber \\
\end{align}
Using $\ref{usefulDid2}$ on the first $\bar{D}$-function of the last three expressions above, gives
\begin{align}
\bar{D}_{k+2 \ k+2 \ n-k+1 \ n+k+1}&=-\bar{D}_{k+2 \ k+2 \ n-k \ n+k+2}-\bar{D}_{k+1 \ k+2 \ n-k+1 \ n+k+2}+n\bar{D}_{k+1 \ k+2 \ n-k \ n+k+1} \nonumber \\
v\bar{D}_{k+2 \ k+2 \ n-k+1 \ n+k+1}&=-\bar{D}_{k+2 \ k+2 \ n-k \ n+k+2}-\bar{D}_{k+2 \ k+1 \ n-k+1 \ n+k+2}+n\bar{D}_{k+2 \ k+1 \ n-k \ n+k+1} \nonumber \\
u\bar{D}_{k+2 \ k+2 \ n-k+1 \ n+k+1}&=-\bar{D}_{k+2 \ k+1 \ n-k+1 \ n+k+2}-\bar{D}_{k+1 \ k+2 \ n-k+1 \ n+k+2}+(2k+1)\bar{D}_{k+1 \ k+1 \ n-k+1 \ n+k+1} 
\nonumber 
\end{align}
Putting everything together, we obtain
\begin{align}
\frac{u}{2k+3}\Big[ - (k(k+1)+n(k+2))(\bar{D}_{k+1 \ k+2 \ n-k \ n+k+1}+\bar{D}_{k+2 \ k+1 \  n-k  \ n+k+1}) \nonumber \\
+(k+2)(2k+1)\bar{D}_{k+1 \ k+1 \ n-k+1 \ n+k+1} \Big]+2u\bar{D}_{k+2 \ k+2 \ n-k \ n+k+2}
\end{align}
The terms in square brackets have $\Sigma=n+k+2$. Noting that
\begin{align}
&\bar{D}_{k+1 \  k+2 \ n-k \ n+k +1}+\bar{D}_{k+2 \  k+1 \ n-k \ n+k +1}=-\bar{D}_{k+1 \ k+1 \ n-k+1 \ n+k+1}+(n+k)\bar{D}_{k+1\ k+1 \ n-k \ n+k}  \nonumber \\
&u\bar{D}_{k+2 \ k+2 \ n-k \ n+k}=\bar{D}_{k+1 \ k+1 \ n-k+1 \ n+k+1}-(n-k-1)\bar{D}_{k+1 \ k+1 \ n-k \ n+k}
\end{align}
and adding the terms in the fourth line of (\ref{along}) we can see that all $\bar{D}$-functions with $\Sigma=n+k+2$ and $\Sigma=n+k+1$ cancel between themselves. Hence (\ref{along}) is given by
\begin{equation}
\tilde{a}(u,v)=2u\bar{D}_{k+2 \ k+2 \ n-k \ n+k+2}
\end{equation}
We may similarly show that the other expressions in (\ref{allshort}) are compatible with (\ref{along})-(\ref{dlong}). Let us now focus on $\tilde{b}_{1}(u,v)$. Take $\Delta_{4}=n+k+2$ in (\ref{combidelta}), which gives for appropriate choices of $\Delta_{1}$, $\Delta_{2}$ and $\Delta_{3}$
\begin{align}
&\bar{D}_{k+3 \ k+1 \ n-k \ n+k+2}+u \bar{D}_{k+3 \ k+2 \ n-k-1 \ n+k+2}+v \bar{D}_{k+2 \ k+2 \ n-k \ n+k+2}=\Gamma(k+2)\Gamma(k+1)\Gamma(n-k-1) \nonumber \\
&\bar{D}_{k+2 \ k+1 \ n-k+1 \ n+k+2}+u \bar{D}_{k+2 \ k+2 \ n-k \ n+k+2}+v \bar{D}_{k+1 \ k+2 \ n-k +1\ n+k+2}=\Gamma(k+1)\Gamma(k+1)\Gamma(n-k) \nonumber \\
&\bar{D}_{k+2 \ k+2 \ n-k \ n+k+2}+u \bar{D}_{k+2 \ k+3 \ n-k-1 \ n+k+2}+v \bar{D}_{k+1 \ k+3 \ n-k \ n+k+2}=\Gamma(k+2)\Gamma(k+1)\Gamma(n-k-1) 
\end{align}
Using these expressions together with (\ref{usefulDid1}), we can evaluate
\begin{align}
(v-u-1)\bar{D}_{k+2 \ k+2 \ n-k \ n+k+2}=&-\Gamma(k+1)^{2}\Gamma(n-k)+n \left( \bar{D}_{k+2 \ k+1 \ n-k-1 \ n+k+2}-\bar{D}_{k+1 \ k+2 \ n-k-1 \ n+k+2}\right) \nonumber \\
&+2 \bar{D}_{k+1 \ k+2 \ n-k \ n+k+3}-2(k+1)\bar{D}_{k+1 \ k+1 \ n-k \ n+k+2}
\label{vucomb}
\end{align}
Also applying (\ref{usefulDid1}) to $\bar{D}_{k+1 \ k+2 \ n-k \ n+k+3}$ we get
\begin{align}
\bar{D}_{k+1 \ k+2 \ n-k \ n+k+3}&=u \bar{D}_{k+3 \ k+2 \ n-k+1 \ n+k+1}+(n-k-1)\bar{D}_{k+2 \ k+1 \ n-k \ n+k+1}
\nonumber \\
&+(k+1)\bar{D}_{k+1 \ k+1 \ n-k \ n+k+2}
\end{align}
so combining this result with (\ref{vucomb}), we can write
\begin{align}
&4u^{2}\bar{D}_{k+3 \ k+2 \ n-k \ n+k+1}=2u\Gamma(k+1)^{2}\Gamma(n-k)+2u(v-u-1)\bar{D}_{k+2 \ k+2 \ n-k \ n+k+2} \nonumber \\
&-4(n-k-1)u\bar{D}_{k+2 \ k+1 \ n-k \ n+k+1}-2nu \left( \bar{D}_{k+2 \ k+1 \ n-k-1 \ n+k+2 }-\bar{D}_{k+1 \ k+2 \ n-k-1 \ n+k+2}\right)
\end{align}
This term is precisely the $\bar{D}$-function that appears in $\tilde{b}_{1}(u,v)$ with $\Sigma={n+k+3}$. All the terms with lower $\Sigma$ appearing above, cancel with the remaining $\bar{D}$-functions in (\ref{b1long}), by means of the identities (\ref{usefulDid1}), (\ref{usefulDid2}) and (\ref{combidelta}), so that we can write
\begin{equation}
\tilde{b}_{1}(u,v)=-2u \Gamma(k+1)\Gamma(k+2)\Gamma(n-k-1)+2u(v-u-1)\bar{D}_{k+2 \ k+2 \ n-k \ n+k+2}
\end{equation}
Similar arguments can be performed for (\ref{b2long}), (\ref{c1long}) and (\ref{c2long}). In the case of $\tilde{d}(u,v)$ we obtain
\begin{align}
\tilde{d}(u,v)&=-\frac{2u^{2}}{v}\Gamma(k+1)^2\Gamma(n-k)+\frac{2u^{2}}{v}(u-v-1)\bar{D}_{k+2 \ k+2 \ n-k \ n+k+2}
\end{align}
\section{Quartic Interactions}
\label{sec:QuarticInt}
We are looking at the quartic diagrams arising in the process
\begin{equation}
\langle \mathcal{O}_{k+2}(\vec{x_{1}}) \mathcal{O}_{k+2}(\vec{x_{2}}) \mathcal{O}_{n-k}(\vec{x_{3}}) \mathcal{O}_{n+k}(\vec{x_{4}})\rangle
\end{equation}
satisfying the sub-sub extremality condition $k_{1}+k_{2}+k_{3}-k_{4}=4$ and $\mathcal{O}_{\Delta}$ are 1/2-BPS operators of $\mathcal{N}=4$ super Yang-Mills theory of conformal weight $\Delta$. The quartic lagrangian reads
\begin{equation}
\mathcal{L}_{4}=\mathcal{L}_{4}^{(0)}+\mathcal{L}_{4}^{(2)}+\mathcal{L}_{4}^{(4)}
\end{equation}
where the supraindex indicates contributions coming from zero, two and four-derivative terms, which are given by
\begin{eqnarray}
\mathcal{L}_{4}&=&\mathcal{L}_{k_{1}k_{2}k_{3}k_{4}}^{(0)I_{1}I_{2}I_{3}I_{4}}s_{k_{1}}^{I_{1}}s_{k_{2}}^{I_{2}}s_{k_{3}}^{I_{3}}s_{k_{4}}^{I_{4}}+
\mathcal{L}_{k_{1}k_{2}k_{3}k_{4}}^{(2)I_{1}I_{2}I_{3}I_{4}}s_{k_{1}}^{I_{1}}D_{\mu}s_{k_{2}}^{I_{2}}s_{k_{3}}^{I_{3}}D^{\mu}s_{k_{4}}^{I_{4}}
\nonumber \\
&+&\mathcal{L}_{k_{1}k_{2}k_{3}k_{4}}^{(4)I_{1}I_{2}I_{3}I_{4}}s_{k_{1}}^{I_{1}}D_{\mu}s_{k_{2}}^{I_{2}}D^{\nu}D_{\nu}(s_{k_{3}}^{I_{3}}D^{\mu}s_{k_{4}}^{I_{4}})
\end{eqnarray}
The explicit form of these terms has been computed in \cite{Arutyunov:1999fb}. 
For our case, two of the $k_{i}$'s are equal to $k+2$. This allows for twelve possible permutations, where the indices $I_{i}$ run over the basis of the representation $[0,k_{i},0]$ which is being summed over. 

We will proceed in similar grounds to those in \cite{Berdichevsky:2007xd,Uruchurtu:2008kp}. We want to express the coefficients as a sum of two terms: one symmetric under exchange of 1-2 and one antisymmetric under the exchange of 1-2.  For the four-derivative couplings there are twelve terms, one for each independent permutation of the $k_{i}$'s. There are two tensor structures entering the expression. These are given by:
\begin{equation}
A^{1234}=A_{1}(\sigma-\tau)+A_{2}(\sigma^{2}-\tau^{2})
\end{equation}
with
\begin{eqnarray}
A_{1}&=&\frac{(2+k) (k-n) (2+2 k-n) (6+2 k+n) (16+4 k (4+k)-n (16+n))}{8192 \sqrt{(k-n) (1+k-n) (-1+k+n) (k+n)} (1+k+n) (2+k+n) (3+k+n)} \nonumber \\
A_{2}&=&-\frac{(k-n) (1+k-n) (2+2 k-n) (6+2 k+n) (52+4 k (7+k)-n (4+n)) }{8192 \sqrt{(k-n) (1+k-n) (-1+k+n) (k+n)} (1+k+n) (2+k+n) (3+k+n)}
\end{eqnarray}
and
\begin{equation}
\Sigma^{1234}=\frac{7 (k+2) (n-k) (n-2-2 k) (n-2) (n+6) (6+2 k+n)\left(\sigma+\tau+\frac{n-k-1}{2(k+2)}(\sigma^{2}+\tau^{2})+\frac{n-k-1}{k+1}\sigma\tau+\frac{k+1}{2(n-k)}\right) }{16384 \sqrt{(n-k) (n-1-k) (n-1+k) (n+k)} (1+k+n) (2+k+n) (3+k+n) (4+k+n)}
\end{equation}
The first tensor is antisymmetric under exchange of $\sigma \leftrightarrow \tau$ whereas the second tensor is symmetric. Using these expressions we can write the four-derivative
terms as:
\begin{align}
\mathcal{L}_{4}^{(4)}&=\frac{1}{2}(A^{1234}+\Sigma^{1234})s^{1}_{k+2}\nabla_{\mu}s^{2}_{k+2}\nabla \cdot \nabla ( s^{3}_{n-k} \nabla^{\mu}s^{4}_{n+k})+\frac{1}{2}
(A^{3412}+\Sigma^{3412})s^{1}_{n-k}\nabla_{\mu}s^{2}_{n+k}\nabla \cdot \nabla ( s^{3}_{k+2} \nabla^{\mu}s^{4}_{k+2}) \nonumber \\
&+\frac{1}{2}(A^{1243}+\Sigma^{1243})s^{1}_{k+2}\nabla_{\mu}s^{2}_{k+2}\nabla \cdot \nabla ( s^{3}_{n+k} \nabla^{\mu}s^{4}_{n-k})+
\frac{1}{2}(A^{3421}+\Sigma^{3421})s^{1}_{n+k}\nabla_{\mu}s^{2}_{n-k}\nabla \cdot \nabla ( s^{3}_{k+2} \nabla^{\mu}s^{4}_{k+2}) \nonumber \\
&-\frac{1}{2}(A^{1324}+\Sigma^{1324})s^{1}_{k+2}\nabla_{\mu}s^{2}_{n-k}\nabla \cdot \nabla ( s^{3}_{k+2} \nabla^{\mu}s^{4}_{n+k})
-\frac{1}{2}(A^{2413}+\Sigma^{2413})s^{1}_{n-k}\nabla_{\mu}s^{2}_{k+2}\nabla \cdot \nabla ( s^{3}_{n+k} \nabla^{\mu}s^{4}_{k+2}) \nonumber \\
&-\frac{1}{2}(A^{1342}+\Sigma^{1342})s^{1}_{k+2}\nabla_{\mu}s^{2}_{n+k}\nabla \cdot \nabla ( s^{3}_{k+2} \nabla^{\mu}s^{4}_{k+2})
-\frac{1}{2}(A^{2431}+\Sigma^{2431})s^{1}_{n+k}\nabla_{\mu}s^{2}_{k+2}\nabla \cdot \nabla ( s^{3}_{n-k} \nabla^{\mu}s^{4}_{k+2}) \nonumber \\
&+\frac{1}{2}A^{1432}s^{1}_{k+2}\nabla_{\mu}s^{2}_{n-k}\nabla \cdot \nabla ( s^{3}_{n+k} \nabla^{\mu}s^{4}_{k+2})+
\frac{1}{2}A^{2341}s^{1}_{n-k}\nabla_{\mu}s^{2}_{k+2}\nabla \cdot \nabla ( s^{3}_{k+2} \nabla^{\mu}s^{4}_{n+k}) \nonumber \\
&+\frac{1}{2}A^{1423}s^{1}_{k+2}\nabla_{\mu}s^{2}_{n+k}\nabla \cdot \nabla ( s^{3}_{n-k} \nabla^{\mu}s^{4}_{k+2})+
\frac{1}{2}A^{2314}s^{1}_{n+k}\nabla_{\mu}s^{2}_{k+2}\nabla \cdot \nabla ( s^{3}_{k+2} \nabla^{\mu}s^{4}_{n-k}) \nonumber 
\end{align}
we use the identity
\begin{equation}
s^{1}_{k_{1}}\nabla_{\mu}s^{2}_{k_{2}}\nabla \cdot \nabla ( s^{3}_{k_{3}} \nabla^{\mu}s^{4}_{k_{4}})=(m_{k_{3}}^{2}+m_{k_{4}}^{2}-4)s_{k_{1}}^{1}\nabla_{\mu}s_{k_{2}}^{2}s_{k_{3}}^{3}\nabla^{\nu}s_{k_{4}}
^{4}+2s_{k_{1}}^{2}\nabla_{\mu}s_{k_{2}}^{2}\nabla_{\nu}s_{k_{3}}^{3}\nabla^{\nu}\nabla^{\mu}s_{k_{4}}^{4}
\end{equation}
and we use the symmetry properties of the tensors. The previous expression becomes
\begin{align}
\mathcal{L}_{4}^{(4)}&= (A^{1234}+\Sigma^{1234})\left(m_{k+2}^2+\frac{m_{n-k}^2+m_{n+k}^2}{2}-4\right)s_{k+2}^{1}\nabla_{\mu}s_{k+2}^{2}
s_{n-k}^{3}\nabla^{\mu}s_{n+k}^{4} \nonumber \\
&+ (A^{1243}+\Sigma^{1243})\left(m_{k+2}^2+\frac{m_{n-k}^2+m_{n+k}^2}{2}-4\right)s_{k+2}^{1}\nabla_{\mu}s_{k+2}^{2}
s_{n+k}^{3}\nabla^{\mu}s_{n-k}^{4} \nonumber \\
&-\Sigma^{1234}\left(m_{k+2}^2+\frac{m_{n-k}^2+m_{n+k}^2}{2}-4\right)(s_{k+2}^{1}s_{k+2}^{2}\nabla_{\mu}s_{n-k}^3\nabla^{\mu}s_{n+k}^4+
s_{n-k}^{1}s_{n+k}^{2}\nabla_{\mu}s_{k+2}^3\nabla^{\mu}s_{k+2}^4) \nonumber \\
&+A^{1243}\left(m_{k+2}^2+\frac{m_{n-k}^2+m_{n+k}^2}{2}-4\right)s_{k+2}^{1}\nabla_{\mu}s_{k+2}^{2}s_{n+k}^{3}\nabla^{\mu}s_{n-k}^{4}
\nonumber \\
&+A^{1234}\left(m_{k+2}^2+\frac{m_{n-k}^2+m_{n+k}^2}{2}-4\right)s_{k+2}^{1}\nabla_{\mu}s_{k+2}^{2}s_{n-k}^{3}\nabla^{\mu}s_{n+k}^{4}
\nonumber
\end{align}
Notice that now all four-derivative terms cancel and the only surviving terms involve at most two derivatives. Integrating by parts the last two terms and simplifying using the symmetries:
\begin{align}
\mathcal{L}^{(4)}_4&= \Sigma^{1234} \left(m_{k+2}^2+\frac{m_{n-k}^2+m_{n+k}^2}{2}-4\right) \left[ s_{k+2}^{1}\nabla_{\mu}s_{k+2}^2 s_{n-k}^{3} \nabla^{\mu}s_{n+k}^{4}+s_{k+2}^{1}\nabla_{\mu}s_{k+2}^2 s_{n+k}^{3} \nabla^{\mu}s_{n-k}^{4}\right. \nonumber \\
&-\left. s_{k+2}^{1}s_{k+2}^{2}\nabla_{\mu}s_{n-k}^{3}\nabla^{\mu}s_{n+k}^{4}-s_{n-k}^{1}s_{n+k}^{2}\nabla_{\mu}s_{k+2}^{3}\nabla^{\mu}s_{k+2}^{4}\right] \nonumber
\end{align}
As in previous cases in the literature, it is possible to simplify this expression further by using the identity:
\begin{equation}
\Omega^{1234}s_{k+2}^{1}\nabla_{\mu}s_{x}^{3} s_{k+2}^{2} \nabla^{\mu}s_{y}^4 = -(\Omega^{1243}+\Omega^{1234})s_{k+2}^{1}\nabla_{\mu}s_{k+2}^{2} s_{x}^{3} \nabla^{\mu}s_{y}^{4}-m_{y}^2 \Omega^{1234}s_{k+2}^{1}s_{k+2}^{2}s_{x}^{3}s_{y}^{4}
\label{idintparts}
\end{equation}
Treating the terms with $(n-k)$ and $(n+k)$ on equal footing, it is possible to arrive to the final expression:
\begin{align}
\mathcal{L}_{4}^{(4)}&= \Sigma^{1234}\left(m_{k+2}^2+\frac{m_{n-k}^2+m_{n+k}^2}{2}-4\right)\left[ 3s_{k+2}^{1}\nabla_{\mu}s_{k+2}^{2}s_{n-k}^{3}\nabla^{\mu} s_{n+k}^{4} \right. \nonumber \\
&+\left.3s_{k+2}^{1}\nabla_{\mu}s_{k+2}^{2}s_{n+k}^{3}\nabla^{\mu} s_{n-k}^{4}+ \left(m_{k+2}^2+\frac{m_{n-k}^2+m_{n+k}^2}{2}\right)s_{k+2}^{1}s_{k+2}^{2}s_{n-k}^{3}s_{n+k}^{4}\right] 
\label{four4term}
\end{align}
We verify again that the four-derivative couplings vanish and that the lagrangian relevant to the computation is of $\sigma$-model type. This gives further evidence that the complete fourth order Lagrangian may share this feature. 

We now present the two-derivative couplings contribution to the quartic lagrangian. The procedure is similar, so that the general structure is of the form
\begin{align}
\mathcal{L}_{4}^{(2)}&=\frac{1}{2}B_{1}^{1234}(s_{k+2}^{1}\nabla_{\mu}s_{k+2}^{2}s_{n-k}^{3}\nabla^{\mu}s_{n+k}^{4}
+s_{k+2}^{1}\nabla_{\mu}s_{k+2}^{2}s_{n+k}^{3}\nabla^{\mu}s_{n-k}^{4}) 
\nonumber \\
&+ \frac{1}{2} B_{2}^{1234}\left(s_{k+2}^{1}s_{k+2}^{2}\nabla_{\mu}s_{n-k}^{3}\nabla^{\mu}s_{n+k}^{4}+\nabla_{\mu}s_{k+2}^{1}\nabla^{\mu}s_{k+2}^{2}s_{n-k}^{3}\nabla^{\mu}s_{n+k}^{4}\right. \nonumber \\
&+\left. s_{k+2}^{1}s_{k+2}^{2}\nabla_{\mu}s_{n+k}^{3}\nabla^{\mu}s_{n-k}^{4}+\nabla_{\mu}s_{k+2}^{1}\nabla^{\mu}s_{k+2}^{2}s_{n+k}^{3}\nabla^{\mu}s_{n-k}^{4} \right)
\end{align}
Making use once more of the identity (\ref{idintparts}), one can rewrite this expression as:
\begin{align}
\mathcal{L}_{4}^{(2)}&=\frac{1}{2}\tilde{B}_{1}^{1234}(s_{k+2}^{1}\nabla_{\mu}s_{k+2}^{2}s_{n-k}^{3}\nabla^{\mu}s_{n+k}^{4}+s_{k+2}^{1}\nabla_{\mu}s_{k+2}^{2}s_{n+k}^{3}\nabla^{\mu}s_{n-k}^{4}) \nonumber \\
&-\frac{1}{2}(m_{n+k}^{2}+m_{k+2}^{2})B_{2}^{1234}s_{k+2}^{1}s_{k+2}^{2}s_{n-k}^{3}s_{n+k}^{4}-\frac{1}{2}(m_{n-k}^{2}+m_{k+2}^{2})B_{2}^{1243}s_{k+2}^{1}s_{k+2}^{2}s_{n+k}^{3}s_{n-k}^{4} \nonumber \\
\label{two4term}
\end{align}
where 
\begin{equation}
\tilde{B}_{1}^{1234}=B_{1}^{1234}-2B_{2}^{1234}-2B_{2}^{1243} 
\end{equation}
Finally we write down the contribution from the non-derivative terms. Using the symmetry $1 \leftrightarrow 2$, one gets:
\begin{equation}
\mathcal{L}_{4}^{(0)}=\frac{1}{2}C_{1}^{1234}(s^{1}_{k+2}s^{2}_{k+2}s^{3}_{n-k}s^{4}_{n+k}+s^{1}_{k+2}s^{2}_{k+2}s^{3}_{n+k}s^{4}_{n-k})
\label{zero4term}
\end{equation}
Combining together equations (\ref{four4term}), (\ref{two4term}) and (\ref{zero4term}), the contribution to the process coming from the quartic lagrangian reduces to:
\begin{equation}
\mathcal{L}_{4}=\tilde{\mathcal{L}}_{4}^{(0)}+\tilde{\mathcal{L}}_{4}^{(2)}
\end{equation}
where
\begin{align}
\tilde{\mathcal{L}}_{4}^{(0)}= \frac{1}{2}\left[ C_{1}^{1234}-\frac{1}{2}(m_{n+k}^{2}+m_{n-k}^{2}+2m_{k+2}^{2})B_{2}^{1234} \right. & +\Sigma^{1234}(m_{k+2}^{2}+\frac{1}{2}m_{n+k}^{2}+\frac{1}{2}m_{n-k}^{2}-4) 
\nonumber \\
& \left. \times(m_{k+2}^{2}+\frac{1}{2}m_{n+k}^{2}+\frac{1}{2}m_{n-k}^{2})\right]s_{k+2}^{1}s_{k+2}^{2}s_{n-k}^{3}s_{n+k}^{4}
\nonumber \\
\end{align}
and
\begin{align}
\tilde{\mathcal{L}}_{4}^{(2)}=\left[3\left( m_{k+2}^{2} + \frac{m_{n-k}^{2}+m_{n+k}^{2}}{2}-4\right)\Sigma^{1234}+\frac{1}{2}\tilde{B}_{1}^{1234}\right]&\left[s_{k+2}^{1}\nabla_{\mu}s_{k+2}^{2}s_{n-k}^{3}\nabla^{\mu}s_{n+k}^{4}\right. \nonumber \\
&+\left.s_{k+2}^{1}\nabla_{\mu}s_{k+2}^{2}s_{n+k}^{3}\nabla^{\mu}s_{n-k}^{4}\right] \nonumber
\end{align}
so after substituting the various quantities and grouping terms
\begin{align}
\mathcal{L}_{4}=\frac{1}{16}(k+1)(k+2)\sqrt{\frac{(n-k)(n-k-1)}{(n+k)(n+k-1)}}&\left[A_{2}^{1234}(s_{k+2}^{1}\nabla_{\mu}s_{k+2}^{2}s_{n-k}^{3}\nabla^{\mu}s_{n+k}^{4}+s_{k+2}^{1}\nabla_{\mu}s_{k+2}^{2}s_{n+k}^{3}\nabla^{\mu}s_{n-k}^{4})\right. \nonumber \\
+&\left.A_{0}^{1234}s_{k+2}^{1}s_{k+2}^{2}s_{n-k}^{3}s_{n+k}^{4}\right] \nonumber \\
\label{4dresult}
\end{align}
where
\begin{align}
A_{2}&= -\tau+\sigma\tau \nonumber \\
A_{0}&=-(k+1)(n+k)+(4k^{2}+k(2+7n)+n(n+6))\sigma+n(3n+k-4)\sigma\tau - n(n+k)\sigma^{2}
\end{align}
%
\section{Exchange Integrals}
\label{sec:exintegrals}
The various method to evaluate exchange diagrams have been thoroughly discussed in \cite{D'Hoker:1999ni, Arutyunov:2002fh,Berdichevsky:2007xd}. The basic idea is to use the underlying symmetries of AdS space to write down an ansatz for the $z$-integral, and then use the Green function equation to determine the explicit functional dependence. As usual, we work in Euclidean $AdS_{d+1}$.
\subsection{Scalar Exchanges}
The scalar exchange integrals have been computed in \cite{D'Hoker:1999ni}. For our case, we only need to consider exchanges of chiral
primaries of weights $2k+2$ and $n$, for the $s$ and $t$ channels, respectively. The generic exchange integral has
the form
\begin{equation}
A(w,\vec{x}_{1},\vec{x}_{2})=\int [dz] G_{\Delta}(z,w)\tilde{K}_{\Delta_{1}}(z,\vec{x}_{1})\tilde{K}_{\Delta_{2}}(z,\vec{x}_{2})
\label{scalarex}
\end{equation}
where $\Delta$ is the conformal weight of the exchanged scalar and $\tilde{K}_{\Delta}(z,x)$ is the unit normalized bulk-to-boundary scalar propagator. The exchange integral can be evaluated by inverting and by making an ansatz for the integrated Green function based on Poincair\'{e} symmetries. For the cases here considered, we quote the results. In the $s$-channel, $\Delta_{1}=\Delta_{2}=k+2$, $\Delta=2k+2$ and $m_{k+2}^2=(k+2)(k-2)$. Hence (\ref{scalarex}) becomes
\begin{equation}
A(w,\vec{x}_{1},\vec{x}_{2})=\frac{1}{4(k+1)^2}|\vec{x}_{12}|^{-2}\tilde{K}_{k+1}(w,\vec{x}_{1})\tilde{K}_{k+1}(w,\vec{x}_{2})
\end{equation}
In the $t$-channel, $\Delta_{1}=k+2$, $\Delta_{3}=n-k$, $\Delta=n$ and $m_{n-k}^{2}=(n-k)(n-k-4)$. The $z$-integral (\ref{scalarex})  gives
\begin{equation}
A(w,\vec{x}_{1},\vec{x}_{3})=\frac{1}{4(k+1)(n-k-1)}|\vec{x}_{13}|^{-2}\tilde{K}_{k+1}(w,\vec{x}_{1})\tilde{K}_{n-k-1}(w,\vec{x}_{3})
\end{equation}
\subsection{Vector Exchanges}
The $z$-integrals for massless and massive vector exchanges have been computed before \cite{D'Hoker:1999ni}. We will just use the results and adapt them to our case. One is interested in diagrams of the form
\begin{equation}
A_{\mu}(w,\vec{x}_{1},\vec{x}_{2})=\int [dz] G_{\mu\nu'}(z,w)g^{\nu'\rho'}(z)\tilde{K}_{\Delta_{1}}(z,\vec{x}_{1})
\frac{\buildrel\leftrightarrow\over\partial}{\partial z_{\rho'}}\tilde{K}_{\Delta_{2}}(z,\vec{x}_{2})
\label{vectorex}
\end{equation}
with $\tilde{K}_{\Delta}(z,\vec{x})$ as before. The propagator once more, transforms as a bitensor under inversion, and it is possible
to proceed in the same lines as in the scalar exchange case. We refer the reader to the formulas in \cite{D'Hoker:1999ni} that determine the solutions. For the $s$-channel, $\Delta_{1}=\Delta_{2}=k+2$ and $m^2=4k(k+1)$ so that 
(\ref{vectorex}) becomes
\begin{align}
A_{\mu}(w,\vec{x}_{1},\vec{x}_{2})
=-\frac{1}{2(k+1)}\frac{1}{|\vec{x}_{12}|^2}&\left\{\frac{(w-\vec{x}_2)_{\mu}}{w_0}\tilde{K}_{k+1}(w,\vec{x}_{1})\tilde{K}_{k+2}(w,\vec{x}_2)\right.
\nonumber \\
&\left.-\frac{(w-\vec{x}_1)_{\mu}}{w_0}\tilde{K}_{k+2}(w,\vec{x}_{1})\tilde{K}_{k+1}(w,\vec{x}_2)
\right\} 
\end{align}
For the $t$-channel, $\Delta_{1}=k+2$, $\Delta_{3}=n-k$ and $m^2=n(n-2)$. The exchange integral evaluates to:
\begin{align}
A_{\mu}(w,\vec{x}_{1},\vec{x}_{3})=\frac{1}{|\vec{x}_{13}|^2}&\left\{ \left(\frac{a_{n-k-1}+2b_{n-k-1}}{2(k+1)}\right)D_{\mu}\tilde{K}_{k+1}(w,\vec{x}_{1})\tilde{K}_{n-k-1}(w,\vec{x}_{3})\right.
\nonumber \\
&\left.-\frac{a_{n-k-1}}{2(n-k-1)}D_{\mu}\tilde{K}_{n-k-1}(w,\vec{x}_{3})\tilde{K}_{k+1}(w,\vec{x}_{1})\right\} \nonumber \\
\end{align}
and in this case, 
\begin{equation}
a_{n-k-1}=-\frac{1}{n} \qquad \qquad b_{n-k-1}=0
\end{equation}

\subsection{Tensor Exchanges}
We now turn to the tensor exchanges. Again, all the ingredients to carry out this computation can be found in the 
literature \cite{D'Hoker:1999ni,Berdichevsky:2007xd}. One needs to compute the $z$-integral
\begin{equation}
A_{\mu\nu}(w,\vec{x}_{1},\vec{x}_{2})=\int [dz] G_{\mu\nu\mu'\nu'}(z,w)T^{\mu'\nu'}(z,\vec{x}_{1},\vec{x}_{2})
\end{equation}
with the tensor $T^{\mu\nu}(z,\vec{x}_{1},\vec{x}_{2})$ being of the form
\begin{eqnarray}
T^{\mu\nu}(w,\vec{x}_{1},\vec{x}_{2})&=&\nabla^{(\mu}\tilde{K}_{\Delta_{1}}(z,\vec{x}_{1})\nabla^{\nu)}\tilde{K}_{\Delta_{2}}(z,\vec{x}_{2})
-\frac{1}{2}g^{\mu\nu}\left(\nabla^{\rho}\tilde{K}_{\Delta_{1}}(z,\vec{x}_{1})\nabla_{\rho}\tilde{K}_{\Delta_{2}}(z,\vec{x}_{2}))\right.
\nonumber \\
&+&\left.\frac{1}{2}(m_{\Delta_{1}}^{2}+m_{\Delta_{2}}^{2}-k(k+4))\tilde{K}_{\Delta_{1}}(z,\vec{x}_{1})\tilde{K}_{\Delta_{2}}(z,\vec{x}_{2})\right)
\end{eqnarray}
where $m_{\Delta}^2=\Delta(\Delta-4)$ and $k$ is the weight of the exchanged tensor, which for our case can be either $2k$ (s-channel) or $n-2$ (t-channel). 

In the $s$-channel amplitude, $\Delta_{1}=\Delta_{2}=k+2$, $m_{\Delta_{1}}^2=m_{\Delta_{2}}^2=k^2-4$ so using manipulations such as the ones presented in \cite{Arutyunov:2002fh} and \cite{Berdichevsky:2007xd}, one can obtain the result
\begin{equation}
A_{\mu\nu}(w,\vec{x}_1,\vec{x}_2)=-\frac{1}{\vert \vec{x}_{12} \vert^{2(k+1)}} \left( \frac{2k+4}{2k+3}\right)Q_{\mu}Q_{\nu} \tilde{K}_{k+1}(w,\vec{x}_1) \tilde{K}_{k+1}(w,\vec{x}_2)
\end{equation}
and for the $t$-channel amplitude, $\Delta_1=k+2$, $\Delta_{3}=n-k$, $\Delta_{13}=2k-n+2$
\begin{equation}
A_{\mu\nu}(w,\vec{x}_{1},\vec{x}_{3})=-\frac{4(k+2)(n-k)}{(n+1)(n+2)}\frac{1}{|\vec{x}_{13}|^2}Q_{\mu}Q_{\nu}\tilde{K}_{k+1}(w,\vec{x}_{3})\tilde{K}_{n-k-1}(w,\vec{x}_{1})
\end{equation}
Here
\begin{equation}
Q_{\mu}=\frac{(w-\vec{x}_{i})_{\mu}}{(w-\vec{x}_{i})^2}-\frac{(w-\vec{x}_{1})_{\mu}}{(w-\vec{x}_{1})^2}
\end{equation}
with $i=2,3$ in the $s$ and $t$-channels respectively.
\section*{Acknowledgements}
I am very grateful to Hugh Osborn for extensive discussions and for sharing his many notes on the subject. His input to this project has proven to be invaluable. I am also thankful for his careful revision of the manuscript. I am indebted to Arkady Tseytlin for discussions and comments. Finally, I'd like to thank Leon Berdichevsky for initial collaboration on this project, and in particular for helping determine the normalisation constants coming from the integrals on $S^5$. 

This work has been supported by a STFC postdoctoral fellowship.

\bibliographystyle{JHEP}
\bibliography{notesref}

\end{document}